%% file: pldi26-differential-privacy.tex
\pdfoutput=1
\RequirePackage{etoolbox}
\newbool{fullversion}
\booltrue{fullversion}
\newbool{arxivversion}
\booltrue{arxivversion}
\RequirePackage{etoolbox}

\providebool{fullversion}
\providebool{arxivversion}

\newbool{needspace}
\booltrue{needspace}

\newbool{anonymous}
\boolfalse{anonymous}

\ifbool{anonymous}{%
\documentclass[acmsmall,review,screen,nonacm,anonymous]{acmart}\settopmatter{printfolios=true,printccs=false,printacmref=false}
}{%
\documentclass[acmsmall,screen,nonacm]{acmart}\settopmatter{printfolios=true,printccs=true,printacmref=false}
}
\usepackage{preamble}
\usepackage{iris}
\usepackage{prob}
\usepackage{examples}

\ifbool{arxivversion}{%
  \citestyle{acmauthoryear}
}{%
\citestyle{acmnumeric}
}

\ifbool{fullversion}{
  \newcommand{\appref}[1]{\cref{#1}}
  \newcommand{\Appref}[1]{\Cref{#1}}
}{
  \ifbool{anonymous}{
    \newcommand{\appref}[1]{the Appendix}
    \newcommand{\Appref}[1]{the Appendix}
  }{%
    \newcommand{\appref}[1]{the Appendix of \cite{clutch-dp:arxiv}}
    \newcommand{\Appref}[1]{The Appendix of \cite{clutch-dp:arxiv}}
  }%
}

\newcommand{\thetitletext}{Modular Verification of Differential Privacy in Probabilistic Higher-Order Separation Logic}
\ifbool{fullversion}
{\title{\thetitletext{} (Extended Version)}}
{\title{\thetitletext{}}}

\author[P. G. Haselwarter]{Philipp~G. Haselwarter}
\orcid{0000-0003-0198-7751}
\affiliation{
  \institution{Aarhus University}
  \country{Denmark}
}
\email{pgh@cs.au.dk}

\author[A. Aguirre]{Alejandro Aguirre}
\orcid{0000-0001-6746-2734}
\affiliation{
  \institution{Aarhus University}
  \country{Denmark}
}
\email{alejandro@cs.au.dk}

\author[S. O. Gregersen]{Simon Oddershede Gregersen}
\orcid{0000-0001-6045-5232}
\affiliation{
  \institution{New York University}
  \country{USA}
}
\email{s.gregersen@nyu.edu}

 \author[K. H. Li]{Kwing Hei Li}
 \orcid{0000-0002-4124-5720}
 \affiliation{
   \institution{Aarhus University}
   \country{Denmark}
 }
 \email{hei.li@cs.au.dk}

\author[J. Tassarotti]{Joseph Tassarotti}
\orcid{0000-0001-5692-3347}
\affiliation{
  \institution{New York University}
  \country{USA}
}
\email{jt4767@cs.nyu.edu}
\authornote{Also affiliated with Amazon Web Services. This paper does not reflect the views of Amazon Web Services.}

\author[L. Birkedal]{Lars Birkedal}
\orcid{0000-0003-1320-0098}
\affiliation{
  \institution{Aarhus University}
  \country{Denmark}
}
\email{birkedal@cs.au.dk}

\ifbool{false}{
\usepackage{caption}
}{
\usepackage[belowskip=-10pt,aboveskip=4pt]{caption}
}

\makeatletter
\ifbool{needspace}{
\def \mpr@medlineskip {\lineskiplimit=1.2em\lineskip=1.0em plus 0.2em}%
\let \MathparLineskip \mpr@medlineskip
}{}
\makeatother

\ifbool{false}{
  \newcommand{\vsquish}[1]{\vspace{-#1}}
}{
  \newcommand{\vsquish}[1]{}
}

\begin{document}

\begin{abstract}
\input{abstract.txt}
\end{abstract}


\setcopyright{cc}
\setcctype{by-nc-nd}
\acmJournal{PACMPL}
\acmYear{2026} \acmVolume{10} \acmNumber{PLDI} \acmArticle{233}
\acmMonth{6} \acmDOI{10.1145/3808311}

\maketitle

\input{introduction.tex}

\input{preliminaries.tex}

\input{logic.tex}

\input{case-studies}

\input{model}

\input{related-work}
\input{conclusion.tex}

\ifbool{anonymous}
{}%
{%
  \input{epilogue}

}

\clearpage

\ifbool{fullversion}{
  \appendix

\clearpage

\input{appendix.tex}

\FloatBarrier
\clearpage
}{}
\bibliography{refs}

\end{document}

%% file: abstract.txt
Differential privacy is the standard method for privacy-preserving data
analysis. The importance of having strong guarantees on the reliability of
implementations of differentially private algorithms is widely recognized and
has sparked fruitful research on formal methods. However, the design patterns
and language features used in modern DP libraries as well as the classes of
guarantees that the library designers wish to establish often fall outside of
the scope of previous verification approaches.

We introduce a program logic suitable for verifying differentially private
implementations written in complex, general-purpose programming languages. Our
logic has first-class support for reasoning about privacy budgets as a
separation logic resource. The expressiveness of the logic and the target
language allow our approach to handle common programming patterns used in the
implementation of libraries for differential privacy, such as privacy filters
and caching. While previous work has focused on developing guarantees for
programs written in domain-specific languages or for privacy mechanisms in
isolation, our logic can reason modularly about primitives, higher-order
combinators, and interactive algorithms.

We demonstrate the applicability of our approach by implementing a verified
library of differential privacy mechanisms, including an online version of the
Sparse Vector Technique, as well as a privacy filter inspired by the popular
Python library OpenDP, which crucially relies on our ability to handle the
combination of randomization, local state, and higher-order functions. We
demonstrate that our specifications are general and reusable by instantiating
them to verify clients of our library. All of our results have been
foundationally verified in the Rocq Prover.

%% file: introduction.tex
\section{Introduction}
\label{sec:intro}

\newcommand{\prog}{P}
\newcommand{\tocite}[1][]{{\color{red!50!black}{[CITE #1]}}}

Differential privacy~\cite{Dwork:Calibrating:2006, Dwork:Algorithmic:2014} (\DP) is a collection of programming techniques to release aggregate information from a database while providing statistical guarantees about the privacy of individual user data.
\DP has been used widely in government and industrial applications to protect critical personal information (medical, financial, demographic, behavioral).
The correctness of the implementations of \DP is therefore crucial: logic- or implementation-level bugs can lead to catastrophic failure of the promised privacy guarantees.

The importance of having trustworthy implementations of \DP is widely recognized. On the one hand, it has led to significant research in programming language and program verification.
On the other, industrial developments of \DP are routinely accompanied by pen-and-paper proofs of their claimed privacy properties. The OpenDP collaboration\footnote{OpenDP was initiated as a collaboration between Harvard University and Microsoft.} even goes as far as to collect a proof for each element of the library, which are checked by a ``privacy proof review board''.

\DP provides strong \emph{statistical} guarantees that the data contributed by an individual does not influence the result of the analysis by too much, and hence cannot be recovered by observing the outcome.
This is achieved by adding a small amount of random noise to each step of data analysis that could leak private information.
This way, even if the input data changes by a small amount (say, the data of one user in a database), we can expect the output of the noisy program output to be similar, and an observer will not be able to tell whether the change comes from the noise or a change in inputs.
The strength of the privacy guarantee of differentially private (\dipr) programs is governed by a parameter \(\err \in \nnreal\), that controls the overall probability that any user's privacy is compromised; a lower $\err$ means a more private program.

\paragraph{Managing the privacy budget}
A central idea of programming with \DP is to think of the parameter \(\err\) that controls the noise distribution as the ``privacy budget''. To ensure that the whole program \(\prog\), written as a sequence of computations \(\prog = c_0 ; c_1 ; \ldots ; c_n\), is \(\err\)-differentially private (\(\err\)-\dipr), we can ``allocate'' some part \(\err_i\) of our budget to each part \(c_i\) of the program and check that the probability that \(c_i\) compromises privacy is controlled according to \(\err_i\).
So long as the sum of the error terms \(\err_i\) is below the global budget \(\err\), the program as a whole is \(\err\)-\dipr.
This principle is known as the sequential composition theorem of \DP, and it justifies our budget intuition of the privacy parameter: \(\err\) represents a \emph{resource} that can be \emph{split} according to the structure of the program, and is \emph{consumed} by computing noisy results from a database.

The most widely used deployments of \DP are based on \emph{dynamic} techniques for tracking the privacy budget via a trusted library such as OpenDP (written in Python, Rust) or Google's Differential Privacy library (C++, Go, Java). Rather than specifying exactly how much \(\err_i\) each part of a program consumes, the budget consumption of the completed computations \(c_0 ; \ldots ; c_{i-1}\) is tracked, and a runtime check ensures that enough budget for \(c_i\) remains.

\subsection{The Challenges of Verifying \DP Frameworks}
\label{sec:challenges}

Most practical \DP frameworks use ``advanced'' programming language features like higher-order functions (\eg, in the form of classes or function pointers) and dynamically allocated, local state (\eg, via class-private attributes).
This is necessary to enable the \emph{modular} construction of private programs through an API, where a library client does not have to concern themselves with the management of the privacy budget and correct application of noise to the results.
To be able to reason compositionally, privacy proofs of a library API should likewise support \emph{modular specifications}.
We consider three representative challenges that arise from reasoning about libraries for \DP:
\begin{enumerate}[label=(\roman*)]
   \item encapsulating dynamic, fine-grained budget accounting,
   \item interactive or ``online'' data analysis,
   \item budget minimization via caching.
\end{enumerate}
As we shall see shortly, all three of these techniques rely on stateful randomized higher-order functions to implement features that are essential to the modular construction of \dipr programs.
To verify modular API specifications we must thus support higher-order reasoning about local state.

We illustrate the programming patterns via Python code snippets inspired by OpenDP, but the challenges also arise in, \eg, the {C++} implementation of Google's Differential Privacy library.\footnote{See, \eg, \url{https://github.com/google/differential-privacy/blob/0a3b05a65f7ed8de/cc/accounting/accountant.cc\#L46}}
Although the verification approach developed in this paper is for an ML-like core language~(\Cref{sec:prelim-lang}) rather than Python, it can faithfully represent the salient aspects of these Python programs.

\subsubsection{Dynamic budget accounting}
\label{sec:intro-filter}

An essential functionality of these libraries is that they offer \emph{privacy filters} \cite{priv-filters} that encapsulate the intricate reasoning about the privacy budget in a core, trusted API.
A privacy filter tracks the privacy cost that a program has incurred up to the current execution point \(c_0;\ldots;c_{i-1}\), and only executes \(c_{i}\) if enough budget remains.
The example implementation of a simplified \texttt{PrivacyFilter} in \Cref{fig:py-filter} is initialized with some \texttt{budget} and provides a single \texttt{try\_run(cost, f)} method, which only runs \texttt{f} if there is indeed at least enough budget left to cover its \texttt{cost}. So long as each mechanism \texttt{f} correctly reports its budget consumption (\texttt{cost}), the filter will ensure that the total privacy cost will never exceed the global budget.

\begin{figure}[htb]
  \centering
  \hfill
 \begin{minipage}[t]{0.39\textwidth}
\begin{python}
class PrivacyFilter:
  def __init__(self, budget):
    self.budget_left = budget

  def try_run(self, cost, f):
    if self.budget_left < cost :
      return None
    self.budget_left -= cost
    return f()
\end{python}
  \end{minipage}
  \hfill
  \mbox{}

  \caption{PrivacyFilter ensures that no client can exceed the privacy budget.}
  \label{fig:py-filter}
\end{figure}

Centralized dynamic budget management simplifies the privacy analysis of programs and allows for scaling to industrial applications.
Importantly, dynamically computed bounds enable a tighter analysis of the privacy cost compared to static type checks \cite{dfuzz,dpella}.
Recall that the usual sequential composition theorem of \DP requires that all \(\err_i\) be chosen upfront. The adaptive composition theorem \cite{priv-filters} for \DP lifts this restriction and allows the analyst to \emph{adaptively} chose each \(\err_i\) depending on the results of previous (private) computations.
Despite the successes of  type- and program-logic-based analyses of \DP, none of the existing systems can be used to specify and verify the correctness of implementations of privacy filters.

\subsubsection{Interactive data analysis}
\label{sec:intro-interactive}

\emph{Interactivity} is central to real-world data analysis under \DP. Given a new dataset, a data analyst does not just issue a static set of queries. Instead, they may compute some summary statistics to explore what ranges of values or categories of responses are of interest, and construct further queries based on those observations.
This interactive or ``online'' style of analysis is crucial for practical utility~\cite{Dwork:Algorithmic:2014,hands-on-dp,turbo}.
A standard way to represent interactive computations is as streams: given a stream of queries, a private interactive mechanism should produce a stream of results.

The \texttt{AboveThreshold} mechanism in \Cref{fig:py-at} constructs a stream of booleans from a stream of queries as an iterator. The \(n\)-th boolean indicates whether \(q_n\) was above the threshold \(T\), after suitable Laplace noise was added. Crucially, the queries themselves are provided as an iterator rather than as a precomputed list, and \(q_n\) is only evaluated in the \(n\)-th call to \texttt{\_\_next\_\_}.
The \texttt{AboveThreshold} mechanism is \(\err\)-\dipr because the stream stops producing results after the first time \texttt{True} is returned; this is a standard result in \DP but challenging to prove formally because it does \emph{not} follow simply from the sequential composition theorem.

\begin{figure}[htb]
  \centering
  \begin{minipage}[t]{0.51\textwidth}
\begin{python}
class AboveThreshold:
  def __init__(self, eps, T, queries, db):
    self.eps = eps
    self.T = Laplace(T, eps/2)
    self.queries = queries
    self.db = db
    self.halted = False

  def __iter__(self):
    return self
\end{python}
  \end{minipage}
    \begin{minipage}[t]{0.48\textwidth}
\begin{python}

  def __next__(self):
    if self.halted:
      raise StopIteration
    q = next(self.queries)
    v = Laplace(q(self.db), self.eps/4)
    b = (self.T <= v)
    self.halted = b
    return b
\end{python}
    \end{minipage}
\caption{The classic Above Threshold interactive mechanism.}
  \label{fig:py-at}
\end{figure}

Any realistic verification framework for \DP must therefore capture not only isolated, static mechanisms but also their behavior as interactive, stateful processes that maintain internal state and privacy budget across calls.

\subsubsection{Budget minimization via caching}
\label{sec:intro-cache}

Certain results (\eg, the number of non-zero entries) have to be calculated many times over  as part of different analysis passes on a dataset.
This can be wasteful in practical applications, since each private computation consumes some of the privacy budget, even if the same result has already been computed elsewhere in the program.
Caching provides a solution to this problem: if the noisy results of queries are cached, then a repeated query can simply reuse the prior noisy result without incurring any privacy cost.
Recent implementations of privacy frameworks have studied increasingly sophisticated caching strategies for \DP workloads~\cite{pioneer,cachedp,privatesql,turbo}, but even simple implementations of caching via memoization can lead to substantial savings in privacy budget in practical workloads~\cite{pioneer}\cite[Fig.~3]{turbo}.

The implementation of memoization \texttt{mk\_query\_cache} in \Cref{fig:py-cache} locally allocates a \texttt{cache} associated to a mechanism \texttt{add\_noise} and a dataset \texttt{db} and returns a closure \texttt{f} which can be used subsequently to privately evaluate queries on \texttt{db}.
Although it is a simple general-purpose caching mechanism, studying it formally requires reasoning about local state and higher-order functions. This places it outside of the scope of existing systems.

\begin{figure}[htb]
  \centering
  \begin{minipage}[t]{0.5\linewidth}
\begin{python}
def mk_query_cache(add_noise, db):
  cache = {}
  def f(query):
    if query in cache:
      return cache[query]
    v = add_noise(query(db))
    cache[query] = v
    return v
  return f
\end{python}
  \end{minipage}
  \caption{A generic caching mechanism.}
  \label{fig:py-cache}
\end{figure}

\subsection{Formal Guarantees for Differential Privacy}
\label{sec:guarantees}
\DP lends itself well to the analysis via standard PL methods due to its compositional nature. We distinguish two
main classes of approaches.
Type-based approaches \cite{fuzz,deAmorim:Probabilistic:2021,dfuzz,afuzz,Barthe:HigherOrder:2015,dpella,param-sens,lightdp,duet,solo,jazz}
use a static typing discipline to ensure that the programs accepted by the system are \dipr.
These approaches benefit from high degrees of automation.
Most type-based approaches, however, do not handle mutable state. More
generally, the complex nature of the type systems required (dependent-,
linear-, contextual-, or refinement types, or combinations thereof), hinders
their integration with mainstream programming languages.

On the other hand, relational probabilistic Hoare logics such as apRHL \cite{relational_via_prob_coupling,apRHL+,advanced_for_diff_privacy}, are more expressive but require more user effort in the form of manual or interactive proofs.
These methods works particularly well when the program logic is defined with respect to a relatively simple denotational semantics of programs as subdistributions.
This, however, generally restricts the applicability of the method to first-order programs.
HO-RPL~\cite{HORHL} extends the previous approach to support higher-order functions by constructing a more sophisticated semantic model, but general recursion or dynamic allocation and higher-order state are still unsupported.
In a similar vein, some projects verified the privacy of sampling algorithms and simple mechanisms directly in the denotational semantics of programs \cite{sampcert,Sato:Formalization:2025}.
In summary, these approaches work well for the verification of algorithms in isolation, but do not focus on supporting \emph{modular} program verification.

\subsection{Modular Verification for \DP Libraries in \thelogic}
\label{sec:modular-dp}

To address the challenges arising from the verification of \DP libraries, we introduce the relational approximate probabilistic higher-order separation logic \thelogic.
\thelogic supports higher-order functions such as \texttt{PrivacyFilter} and \texttt{mk\_query\_cache} via quantification over specifications, and references storing closures as they occur in the implementation of iterators (\eg, \texttt{AboveThreshold}) via impredicative invariants à la Iris~\cite{irisjournal}.
Building on previous relational separation logics \cite{clutch,approxis}, we internalize the privacy budget of \DP in \thelogic as a first-class separation logic resource.
These \emph{privacy credits} can be tracked in invariants and interact with the heap and with higher-order functions that consume a privacy budget in flexible ways.
We prove that the usual rules for (sequential) relational separation logic~\cite{reloc} are sound in the probabilistic setting.
To reason about privacy of primitives, we prove novel relational sampling rules for the Laplacian; in particular, we internalize a proof technique based on choice couplings~\cite{Albarghouthi:Synthesizing:2017} as a logical rule, which, in particular, is used in the privacy proof of \texttt{AboveThreshold}.
We apply \thelogic to a number of case studies inspired by the challenge problems described above.
Our specifications of, \eg, \texttt{AboveThreshold} are compositional and reusable: we derive the privacy of clients and more complex mechanisms from the specifications of the building blocks, without referring to implementation details.

\paragraph{Contributions}
\label{sec:contributions}
To summarize, we make the following contributions.
\begin{enumerate}
\item \emph{A higher-order separation logic for \DP} which internalizes privacy credits as first-class, composable logical resources and which supports heap allocation and higher-order closures.
\item \emph{New probabilistic sampling rules}, including a Laplacian rule that enables selective recovery of privacy credits,
  which can be used to verify examples whose privacy analysis goes beyond composition theorems.
\item \emph{A library of reusable, abstract specifications} for common \DP primitives (Laplace, AboveThreshold, Sparse Vector Technique, Report-Noisy-Max, Privacy Filters, Caching via Memoization) that cleanly separate mechanism proof from client reasoning.
\item \emph{Client case studies} that highlight both expressiveness and reusability of specifications and that demonstrate that our approach successfully addresses the challenges outlined above (dynamic budget accounting, interactive data analysis, and budget minimization via caching).
\item \emph{Mechanized proofs}: a foundational formalization of the logic and all case studies in the \rocq Prover together with an
  adequacy theorem connecting our logic to standard \DP.
\end{enumerate}

\paragraph{Outline}
In \Cref{sec:preliminaries} we briefly recall the basic notions of \DP and define the \thelang programming language used in the remainder of the paper. In \Cref{sec:program-logic} we define the \thelogic logic and explain how it relates to \DP. In \Cref{sec:case-studies} we illustrate reasoning in \thelogic via a number of case studies addressing the challenges set out in \Cref{sec:challenges}. The soundness of \thelogic is addressed in \Cref{sec:model}. Finally, in \Cref{sec:related-work} we survey related work, before concluding in \Cref{sec:conclusion-future-work}.%
\footnote{%
  \ifbool{arxivversion}{
      This is the extended version of \cite{clutch-dp}, which differs from the published version by the inclusion of the full Appendix.
    }{
      \ifbool{anonymous}{%
          \ifbool{fullversion}{%
              This is the full version of the paper, which includes a full appendix.
            }{
              The full version of this paper including the appendix is available as supplementary material.
            }
        }{%
          \ifbool{fullversion}{%
              This is the full version of \cite{clutch-dp}, which includes a full appendix.
            }{
              The extended version of this paper including the appendix is available as \cite{clutch-dp:arxiv}.
            }
        }
    }
}

%% file: preliminaries.tex
\section{Differential Privacy and Programming Language Preliminaries}
\label{sec:preliminaries}

We briefly recall the elements of probability theory \Cref{sec:prelim-prob} and \DP \Cref{sec:prelim-dp} we need to refer to, and formally define the programming language \Cref{sec:prelim-lang} used throughout the paper.

\subsection{Probability Theory}
\label{sec:prelim-prob}

Since we do not assume a priori that all programs we study terminate, we allow programs to ``lose mass'' on diverging runs and define the operational semantics using probability \emph{sub}-distributions.
\begin{definition}
  A \defemph{discrete subdistribution} (henceforth simply \defemph{distribution}) on a countable set $A$ is a function $\distr : A \ra [0,1]$ such that $\sum_{a \in A}\distr(a) \leq 1$.
  The distributions on $A$ are denoted by $\Distr A$.
\end{definition}
In \Cref{sec:opsem}, we will define the operational semantics of \thelang in terms of the distribution monad.
\begin{lemma}
  The \defemph{discrete distribution monad} induced by $\DDistr$ has operations
	\begin{align*}
		&{}\mret{} \colon A \to \Distr A \qquad &&{}\mbind{}\colon (A \to \Distr B) \to \Distr A \to \Distr B \\
		&{}\mret(a)(a') \eqdef{}
          \begin{cases}
            1 & \text{if } a = a' \\
            0 & \text{otherwise}
          \end{cases} &&{}\mbind(f,\distr)(b) \eqdef{} \sum_{a \in A} \distr(a) \cdot f(a)(b)
	\end{align*}
  We write $(\distr \mbindi f)$ for $\mbind(f, \distr)$.
\end{lemma}

\subsection{Differential Privacy}
\label{sec:prelim-dp}

For background information on \DP see, \eg, \citet{Dwork:Algorithmic:2014} and \citet{hands-on-dp}.

\paragraph{Defining privacy}
The definition of \DP captures the intuition that it is hard to reconstruct information about any individual in a database from the output of a \dipr program.
\begin{definition}\label{def:dp}
  A function \(f : \DB \to \Distr X\) is \((\merr,\aerr)\)-differentially private (short: ``\(f\) is \((\merr,\aerr)\)-\dipr'') if\; \(\pr[f(x)] \pprop \leq e^\err \cdot \pr[f(y)] \pprop + \aerr\)\, for all adjacent \(x,y : \DB\) and all predicates \(\pprop \subseteq X\).
  If \(f\) is \((\err, 0)\)-\dipr we simply say that \(f\) is \(\err\)-\dipr.
\end{definition}
A function is thus \((\err,\aerr)\)-\dipr if it amplifies the probability of \emph{any} observation \(\pprop\) by at most~\(e^\err\), or by more than that with probability at most \(\aerr\).
A small value for \(\err\) and \(\aerr\) thus means strong privacy guarantees.
The definition of \DP is parametrized by a type of databases \(\DB\) and an adjacency relation. \thelogic can work with any database type and adjacency relation, but a common choice is to think of a database as a list of rows where each row is a tuple of a fixed size.
If the type of databases comes with a notion of distance \(\dd[\DB] : \DB \to \nnreal\) we say that two databases \(x, y\) are adjacent if \(\dr[\DB] x y 1\). For instance, if \(\df[\DB] x y\) is the number of rows where \(x\) and \(y\) differ, then adjacency means they only differ in one row.

\paragraph{Adding noise: the Laplace mechanism}

\begin{wrapfigure}{R}{0.4\textwidth}\label{fig:laplace}
  \includegraphics[width=\linewidth]{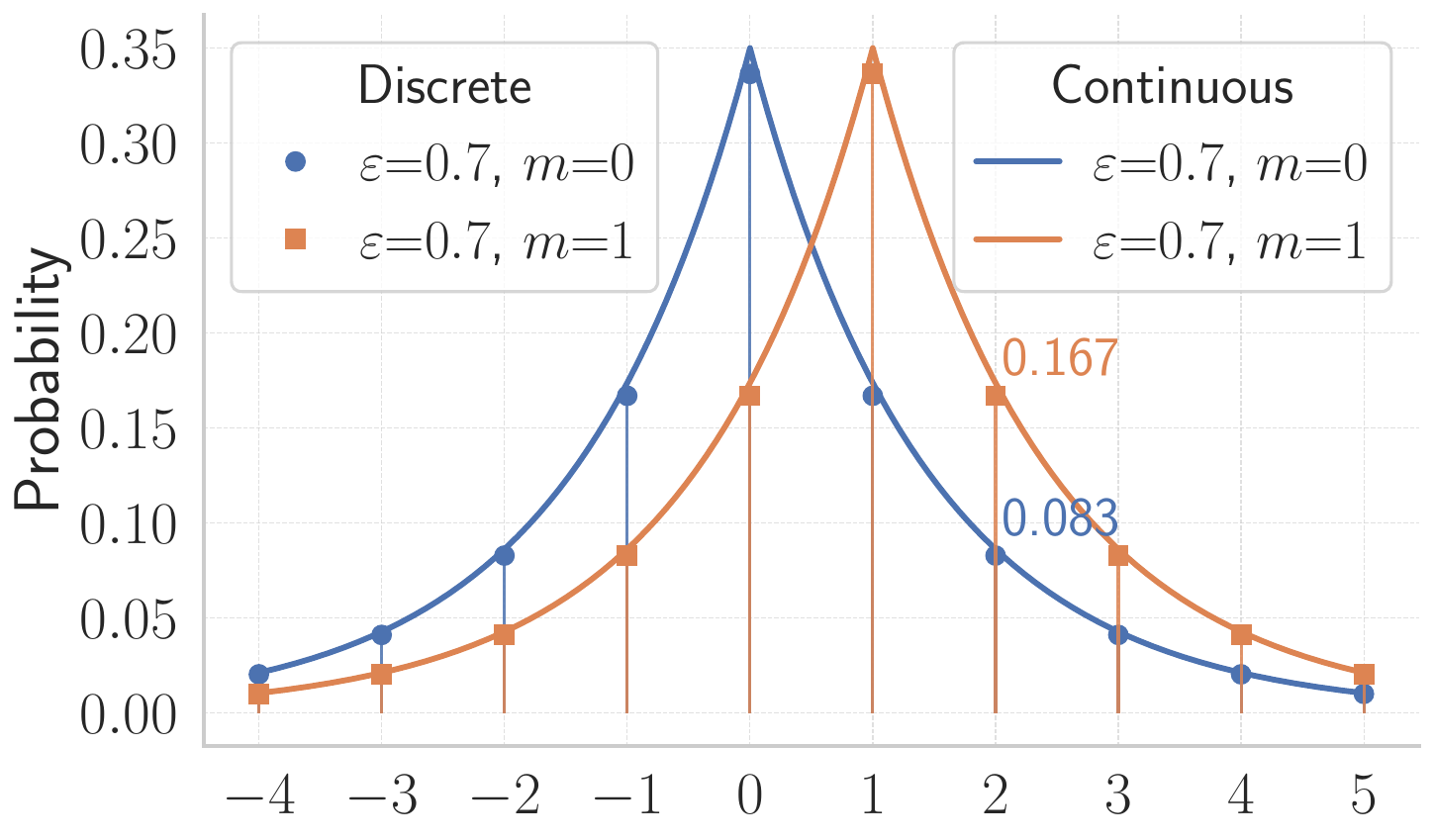}
  \caption{Continuous and discrete Laplacian, with \(\laplacepmf_{0.7}^\lapmean(\val)\) for \(\val{=}2\) and \(m\in \set{0,1}\), demonstrating
   \(0.7\)-\DP: \(\laplacepmf_{0.7}^1(2) \leq e^{0.7} \cdot \laplacepmf_{0.7}^0(2)\)
    .}
\end{wrapfigure}
To prove that any program is \dipr, we need primitives that add random noise.
The prototypical example of a noise mechanism that achieves \DP is the Laplacian distribution. However, since the Laplacian is a continuous distribution on \(\real\), an implementation of a Laplacian sampler would have to work with exact real arithmetic, since implementations using floats lead to well-known privacy bugs~\cite{Mironov:significance:2012}.

We therefore work with the \defemph{discrete Laplacian} (\hyperref[fig:laplace]{Fig. 4}), the distribution on \(\integer\) obtained by discretizing
the continuous Laplacian.
The discrete Laplacian with scale parameter \(\err\) and mean \(\lapmean\) has as probability mass function\footnote{NB: The weight \(W\) is a geometric series over \(\integer\) (hence why \(\laplacepmf_\err\) is also called the (two-sided) \(\err\)-geometric) with the closed form \(\laplacepmf_\err^\lapmean(\val) = \frac{e^\err - 1}{e^\err + 1} \cdot e^{-\err {\cdot} \abs {\val-\lapmean}}\)~\cite{Canonne:discrete:2020}. We adopt the convention that \(\laplacepmf_\err^\lapmean = \mret(\lapmean)\) if \(\err \leq 0\).}~\cite{Inusah:discrete:2006,Hsu:Probabilistic:2017}:
\begin{equation}
  \label{eq:laplace-pmf}
  \laplacepmf_\err^\lapmean(\val) \eqdef \frac{1}{W} \cdot e^{- \err {\cdot} \abs {\val - \lapmean}}
  \qquad
  \ifbool{arxivversion}%
    {\makebox[85mm][l]{where \(\displaystyle W \eqdef \sum_{z \in \integer} e^{- \err \abs z}\)}}%
    {\makebox[10mm][l]{where \(\displaystyle W \eqdef \sum_{z \in \integer} e^{- \err \abs z}\)}}
\end{equation}

Interpreting adjacency on \(\integer\) as being at distance at most~\(\scale\), we have the following result.
\begin{theorem}[\cite{Ghosh:Universally:2012}]\label{thm:laplace-dp}
  \((\Lam \lapmean . \laplacepmf_{\err/\scale}^\lapmean) : \integer \to \Distr \integer\) is \(\err\)-\dipr.
\end{theorem}
As a direct consequence, given any function \(f : \DB \to \integer\) such that \(\abs{f(x) - f(y)} \leq 1\) for adjacent datasets \(x,y : \DB\), the function \((\Lam \lapmean . \laplacepmf_\err^\lapmean) \circ f : \DB \to \Distr \integer\) is \(\err\)-\dipr.

\paragraph{Composing \DP}
Differentially private functions satisfy a number of composition laws that can help structure and simplify the privacy verification of larger programs. %

\begin{lemma}\label{thm:post-comp}
  \DP is \defemph{stable under post-processing}:
  if \(f : A \to \Distr B\) is \((\err,\aerr)\)-\dipr then for any \(g : B \to C\), the function \(\Lam x. (f (x) \mbindi \Lam y . \mret\ g (y)) : A \to \Distr C\) is \((\err,\aerr)\)-\dipr.
\end{lemma}

When two \dipr functions are composed sequentially, their privacy parameters add up. Due to the post-processing property this holds even if the later computations can see the results of earlier computations.
We only show the case for two functions, but the lemma directly generalizes to arbitrary \(k\)-fold composition for \(k \in \nat\).
\begin{lemma}[Sequential Composition]\label{thm:seq-comp}
  Let \(f : \DB \to \Distr B\) be \((\err_1, \aerr_1)\)-\dipr and let \(g\) be a function \(g : \DB \times B \to \Distr C\) such that \((\Lam x . g(x, b)) : \DB \to \Distr C\) is \((\err_2, \aerr_2)\)-\dipr for all \(b \in B\). Then \(\Lam x . (f (x) \mbindi \Lam b . g (x, b))\) is \((\err_1+\err_2, \aerr_1 + \aerr_2)\)-\dipr.
\end{lemma}

Another useful composition law holds for functions which increase the distance between inputs by at most a fixed amount \(\scale\) in the following sense.
\begin{definition}\label{def:sens}
  We say that \(f : A \to B\) is \defemph{\(\scale\)-sensitive} (also: ``\defemph{\(\scale\)-stable}'') if for all \(x,y \in A\), the bound \(\dr[B]{f(x)}{f(y)}{\scale \cdot \df[A] x y}\) holds, where the distances \(\dd[A], \dd[B]\) are taken with respect to a metric space structure on \(A\) and \(B\).
\end{definition}
Note that if \(f:A\to B\) is \(\scale\)-sensitive and \(g : B \to C\) is \(\scaleb\)-sensitive then \(g \circ f\) is \((c{\cdot}d)\)-sensitive.
The following \defemph{metric composition law} then generalizes the remark following \cref{thm:laplace-dp}.
\begin{lemma}\label{thm:metric-comp}
  If \(f : \DB \to \integer\) is \(\scale\)-sensitive then \((\Lam \lapmean . \laplacepmf_{\err/\scale}^\lapmean) \circ f : \DB \to \Distr \integer\) is \(\err\)-\dipr.
\end{lemma}

\subsection{The Language: Randomized ML}
\label{sec:prelim-lang}

The $\thelang{}$ language that we consider is an ML-like language with higher-order recursive functions and higher-order state that we extend with an operator \(\Laplace a\, b\, \lapmean\) that samples from the Laplacian with scale \((a/b)\) and mean \(\lapmean\).
The syntax is defined by the grammar below.
\begin{align*}
  \val, \valB \in \Val \bnfdef{}
  & z \in \integer \ALT
  b \in \bool \ALT
  \TT \ALT
  \loc \in \Loc \ALT
  \Rec \vf \lvar = \expr \ALT
  (\val,\valB) \ALT
  \Inl \val  \ALT
  \Inr \val
  \\
  \expr \in \Expr \bnfdef{}  &
  \val \ALT
  \lvar \ALT
  \Rec \vf \lvar = \expr \ALT
  \expr_1~\expr_2 \ALT
  \expr_1 + \expr_2 \ALT
  \expr_1 - \expr_2 \ALT
  \ldots \ALT
  \If \expr then \expr_1 \Else \expr_2 \ALT
  (\expr_1,\expr_2) \ALT
  \Fst \expr \ALT \ldots \\
  & \Alloc~\expr_1 \ALT
  \deref \expr \ALT
  \expr_1 \gets \expr_2 \ALT
  \Laplace~\expr_1~\expr_2~\expr_3 \ALT
    \ldots \\
  \lctx \in \Ectx \bnfdef{}  &
  -
  \ALT \expr\,\lctx
  \ALT \lctx\,\val
  \ALT \Alloc~\lctx
  \ALT \deref \lctx
  \ALT \expr \gets \lctx
  \ALT \lctx \gets \val
  \ALT \Laplace \expr_1~\expr_2~\lctx
  \ALT \Laplace \expr~\lctx~\val
  \ALT \ldots
  \\
  \state \in \State \eqdef{} & \Loc \fpfn \Val \hspace{2em} \cfg \in \Conf \eqdef{} \Expr \times \State %
\end{align*}
In \thelang, $\Alloc \expr_1$ allocates a new reference containing the value returned by $\expr_1$,
$\deref \expr$ dereferences the location $\expr$ evaluates to, and $\expr_{1} \gets \expr_{2}$ evaluates $\expr_{2}$ and assigns the result to the location that $\expr_{1}$ evaluates to.
We may refer to a recursive function value $\Rec \vf \lvar = \expr$ by its local name $\vf$. The heap is represented as a (partial) finite map from locations to values, and evaluation happens right to left as indicated by the evaluation context grammar \(\Ectx\).

The expression \(\Laplace a\,b\,\lapmean\) samples from the discrete Laplacian with scale \(\err = a / b\) and mean~\(\lapmean\). To avoid unnecessary complications with adding real numbers to the programming language, we require the scale \(\err\) to be a rational number. Formally, in \thelang, \(\Laplace\) takes three integers as input, but to keep our notation free from clutter we will simply write \(\Laplace \err\, \lapmean\) instead of \(\Laplace a\,b\, \lapmean\) with \(\err = a/b\).

\paragraph{Operational Semantics}
\label{sec:opsem}

Program execution is defined by iterating $\stepdistr : \Conf \to \Distr{\Conf}$, where \(\stepdistr(\cfg)\) is the distribution induced by the single step reduction of the configuration~$\cfg$.
The semantics is mostly standard.
We first define head reduction and then lift it to reduction in an evaluation context $K$.
All non-probabilistic constructs reduce deterministically as usual, \eg, %
$\stepdistr((\Fun \vx . \expr)\,\val, \sigma) = \mret(\expr[\val/\vx], \sigma)$.
We write \(e \purestep e'\) if the evaluation is deterministic and holds independently of the state, \eg, \((\Fun \vx . \expr)\,\val \purestep \expr[\val/\vx]\) and \(\Fst (\val_1, \val_2) \purestep \val_1\).
The sampling operator $\Laplace a\,b\,\lapmean$ reduces according to the Laplacian with scale \(a/b\) and mean \(\lapmean\), \ie,
\begin{align}\label{eq:opsem-laplace}
  & \stepdistr(\Laplace a\,b\,\lapmean, \sigma)(\val, \sigma) \eqdef{}
  \begin{cases}
    \laplacepmf_{a/b}^\lapmean(\val) & \text{for } \val \in \integer, \\
    0                        & \text{otherwise}.
  \end{cases}
\end{align}

With the single step reduction $\stepdistr : \Conf \to \Distr{\Conf}$ defined, we next define a step-stratified execution probability $\exec_{n}\colon \Conf \to \Distr{\Val}$ by induction on $n$:
\begin{align*}
	\exec_{0}(\expr, \state)(\val) &\eqdef{}
 \begin{cases}
   1                                           & \text{if}~\expr\in\Val \wedge \expr = \val, \\
   0 & \text{otherwise.}
 \end{cases}\\
	\exec_{n+1}(\expr, \state)(\val) &\eqdef{}
 \begin{cases}
   1                                           & \text{if}~\expr\in\Val \wedge \expr = \val, \\
   \sum_{(\expr',\state')\in \Expr\times\State} \stepdistr(\expr, \state)(\expr',\state') \cdot \exec_{n}(\expr',\state')(\val) & \text{otherwise.}
 \end{cases}
\end{align*}
That is, $\exec_{n}(\expr, \state)(\val)$ is the probability of stepping from the configuration $(\expr, \state)$ to a value $\val$ in at most $n$ steps.
The probability that an execution, starting from configuration $\cfg$, reaches a value $\val$ is taken as the limit of its stratified approximations, which exists by monotonicity and boundedness:
\begin{align*}
 \exec(\cfg)(\val) \eqdef{} \lim_{n \to \infty} \exec_{n}(\cfg)(\val)
\end{align*}

The interpretation of programs as distributions induces a natural notion of \((\err, \aerr)\)-\DP for \thelang programs. Concretely, if \(f \in \Expr\) is a \thelang function then for adjacent databases \(x,y\) it should be the case that for all states \(\state\),
\begin{equation}
  \label{eq:exec-dp}
  \pr[\exec(f\ \inject(x),\ \state)]{\pprop} \leq e^\err \cdot \pr[\exec(f\ \inject(y),\ \state)]{\pprop} + \aerr
\end{equation}
where \(\inject : \DB \to \Val\) embeds the type of databases into \thelang values (we usually omit \(\inject\)).

Note that in particular \(f = (\Fun \vx . \Laplace \err\, \vx)\) is \(\err\)-\dipr in the sense of \eqref{eq:exec-dp} by the definition of the operational semantics~\eqref{eq:opsem-laplace} and \Cref{thm:laplace-dp}.

%% file: logic.tex
\section{Program Logic}
\label{sec:program-logic}

In this section, we introduce the \thelogic logic, the soundness theorem of \thelogic, which connects it to \DP, and the new logical connectives and rules pertaining to the privacy budget reasoning.
Our logic takes inspiration from Approxis~\cite{approxis}, but we remark that the underlying model is different, and that it is designed to support our novel reasoning principles for \DP via multiplicative error credits, rules for the Laplacian, and new composition laws.

\paragraph{The logical connectives} \thelogic is built on top of the Iris separation logic framework~\cite{irisjournal} and inherits many of Iris's logical connectives, a selection of which is shown below.
Most of the propositions are standard, such as separating conjunction $\prop \sep \propB$ and separating implication $\prop \wand \propB$.
\begin{align*}
  \prop,\propB \in \iProp \bnfdef{}
  & \TRUE \ALT \FALSE \ALT \prop \land \propB \ALT \prop \lor \propB \ALT \prop \Ra \propB \ALT
    \All \var . \prop \ALT \Exists \var . \prop \ALT \prop \sep \propB \ALT \prop \wand \propB \ALT \\
  & \progheap{\loc}{\val} \ALT
    \specheap{\loc}{\val} \ALT
    \uptom{\err} \ALT
    \upto{\aerr} \ALT
    \hoareref{\prop}{\expr}{\expr'}{\val\ \val' \ldotp \propB} \ALT
     \ldots
\end{align*}
Since \DP is a relational property, \thelogic is a relational program logic that proves properties about the execution of two programs \(\expr\) and \(\expr'\).
The central logical connective capturing the relationship between \(\expr\) and \(\expr'\) is the \defemph{Hoare quadruple}:
\begin{equation}
  \label{eq:hoare-quad}
  \hoareref{\prop}{\expr}{\expr'}{\val\ \val' \ldotp \propB}
\end{equation}
Intuitively, this quadruple asserts that under the precondition \(\prop\), if
the two programs \(\expr\) and \(\expr'\) evaluate to results \(\val\) and
\(\val'\) respectively, then the postcondition \(\propB\) holds. Since the
pre- and postcondition range over arbitrary \thelogic assertions, they can refer
to the state or contain nested Hoare quadruples, which will be useful to write
specifications about higher-order functions.

Logical assertions about the heap come in two versions, one for \(\expr\) and one for \(\expr'\).
The heap points-to assertion that denotes ownership of location $\loc$ for the left-hand side program~\(\expr\) is written as $\progheap{\loc}{\val}$, and $\specheap{\loc'}{\val'}$ denotes ownership of \(\loc'\) for the right-hand side program~\(\expr'\).

\thelogic defines two kinds of resources that are novel with respect to standard Iris and which are inspired by Approxis~\cite{approxis}.
The proposition \(\uptom \err\) asserts ownership of \(\err\) \defemph{multiplicative privacy credits} and \(\upto \aerr\) similarly asserts ownership of \(\aerr\) \defemph{additive privacy credits}.
These credits are a logical representation of the privacy budget of \DP.
Just as reasoning about the physical state of a program is logically captured by the points-to connective, so is reasoning about the privacy budget \((\err, \aerr)\) logically expressed via privacy credits.\footnote{Contrary to heaps, the privacy budget pertains to a pair of program executions rather than to the left or right program, and hence there is no left- and right-hand version of credits; instead, they are shared.}

\paragraph{Internalizing privacy} We now have all of the ingredients to define \DP in \thelogic.
\begin{definition}\label{def:idp}
  A \thelang function \(\vf\) is \defemph{internally \((\err,\aerr)\)-\dipr} if the following quadruple holds:
	\begin{equation}\label{eq:idp}
		\All \vdb\,\vdb'. \hoareref{\adj{\vdb}{\vdb'} \sep \uptom \err \sep \upto \aerr}{\vf\, \vdb}{\vf\, \vdb'}{\val,\, \val' \ldotp \val = \val'}
	\end{equation}
\end{definition}
As with \cref{def:dp}, this
notion is parametrized by an adjacency relation $\adj{\vdb}{\vdb'}$ on
inputs. For instance, we could represent the databases as
lists, where each element represents one entry, and say two databases are adjacent if
they differ only in one entry (when considered as multisets).
We abbreviate the Hoare quadruple \eqref{eq:idp} as \(\hoaredp\err\aerr\vf\)
when the adjacency relation is clear from context, and simply \(\hoaredppure\err\vf\) for
the case where $\aerr=0$.

{%
  \crefname{equation}{Eq.}{Eqs.}\crefname{definition}{Def.}{Defs.}
  We call \Cref{def:idp} \defemph{internal} \DP since it is a property of the \thelang{} program \(\vf\) stated entirely in terms of the program logic. All reasoning about concrete probabilities is abstracted away via the use of privacy credits.
  In contrast, the usual \defemph{external} definition of \DP (\cref{def:dp}) for arbitrary functions and for \thelang{} programs (\cref{eq:exec-dp}) is given in terms of plain mathematics, probabilities, and the operational semantics.
  Working with internal \DP has two advantages: (1) It uses an expressive program logic, which allows us to reason about complex programs, whereas reasoning directly about the operational semantics would be infeasible. (2) We can ``pay'' some privacy credits to get an equality in the postcondition of \cref{eq:idp}, so that a client of~\(\vf\) can simply assume that the results of  \((\vf\ \vdb)\) and  \((\vf\ \vdb')\) are \emph{identical} instead of having to work with the assumption that they are distributions at distance \((\err,\aerr)\).
}

The precise meaning of Hoare quadruples and privacy credits can be understood through the following theorem which connects \thelogic to \DP in the sense of \cref{eq:exec-dp}.
\begin{theorem}[Soundness]
\label{thm:soundness-dp}
  If\, \(\vf\) is internally \((\err,\aerr)\)-\dipr then \(\vf\) is also externally \((\err,\aerr)\)-\dipr.
\end{theorem}
This theorem is a consequence of the adequacy of the semantic model of \thelogic (see \Cref{sec:model}).

\subsection{Relational Separation Logic Rules}
\label{sec:standard-rules}

The standard rules of relational separation logic as used in, \eg, \cite{reloc} are also available in \thelogic.
For instance, the \rref{Frame} rule enables local reasoning by framing out \(\propC\), and the \rref{Bind} rule allows us to focus on sub-programs in evaluation contexts.
The usual load and store rules for heap locations come in a left- and a right version, requiring ownership of the corresponding points-to connective.

\begin{figure}[h]
  \centering
  \begin{mathpar}
    \small
    \infrule{Frame}
    {\hoareref{\prop}{\expr}{\expr'}{\propB}}
    {\hoareref{\prop \sep \propC}{\expr}{\expr'}{\propB \sep \propC}}
    \quad
    \infrule{Bind}
    {
      \hoareref{\prop}{\expr}{\expr'}{\val\,\val' \ldotp \propC}
      \and
      \All \val\,\val' .
      \hoareref{\propC}{\fillctx \lctx [\val] }{\fillctx \lctx' [\val'] }{\propB}
    }
    {\hoareref{\prop}{\fillctx \lctx[\expr]}{\fillctx \lctx'[\expr']}{\propB}}
    \\
    \infrule{Load-l}
    {\hoareref{\loc \pointsto \valB}{\valB}{\expr'}{\propB}}
    {\hoareref{\loc \pointsto \valB}{\deref \loc}{\expr'}{\propB}}
    \and
    \infrule{Load-r}
    {\hoareref{\loc' \spointsto \valB}{\expr}{\valB}{\propB}}
    {\hoareref{\loc' \spointsto \valB}{\expr}{\deref \loc'}{\propB}}
    \and
    \infrule{Store-l}
    {\hoareref{\loc \pointsto \valB}{\TT}{\expr'}{\propB}}
    {\hoareref{\loc \pointsto \val}{\loc \gets \valB}{\expr'}{\propB}}
  \end{mathpar}
  \caption{Excerpt of the non-probabilistic rules of \thelogic.}
  \label{fig:standard-rules}
\end{figure}

It is worth noting that the rules in \Cref{fig:standard-rules} do \emph{not} mention distributions, despite the fact that the operational semantics of \thelang is probabilistic in general and references can, for instance, store randomly sampled values.
\thelogic thus provides a convenient basis for program verification where reasoning about privacy is integrated as an \emph{orthogonal} feature while preserving the familiar rules of relational higher-order separation logic.

\subsection{Privacy Credit Laws}
\label{sec:logic-priv-cred}

The privacy credit resources logically track two non-negative real numbers \(\err\) and \(\aerr\) corresponding to the privacy budget. In light of this intuition, one would expect that they support laws such as sequential composition (\cref{thm:seq-comp}). Indeed, we can derive an internal version of this law from the primitive rules pertaining to privacy credits show in \Cref{fig:privacy-credits} together with the structural rules for \thelang in \Cref{fig:standard-rules}.

\begin{figure}[h]
  \centering
  \begin{mathpar}
    \uptom {\err_1 + \err_2} \dashv\vdash \uptom {\err_1} \sep  \uptom {\err_2}
    \and
    \upto {\aerr_1 + \aerr_2} \dashv\vdash \upto {\aerr_1} \sep  \upto {\aerr_2}
    \and
    \upto 1 \vdash \FALSE
  \end{mathpar}
  \caption{Privacy credit laws.}
  \label{fig:privacy-credits}
\end{figure}

Just as in \Cref{thm:seq-comp}, both multiplicative and additive privacy credits can be split into their summands\footnote{NB: Both kinds of privacy credit are split into separating conjunctions by using addition, not multiplication, as the resource algebra operation. The name ``multiplicative privacy credit'' for \(\uptom \err\) derives from the multiplicative factor \(e^\err\) in \cshref{def:dp}.} (and recombined).
The last rule states that from 1 additive error credit, we can derive \(\FALSE\) and hence anything. This makes sense if we recall that \(\aerr\) models a bound on an inequality between \emph{probabilities} in the definition of \DP (\cshref{def:dp}), since any probability is bounded by 1, and hence any program is trivially \((\err,1)\)-\dipr.

The fact that the privacy budget is treated like any other separation logic resource and being able to split the budget enables \emph{flexible} reasoning about privacy.
Privacy credits are animated by the rules that govern their interaction with the operational semantics of \thelang via sampling.

\subsection{Rules for Sampling Noise}
\label{sec:sampling-rules}

Privacy credits are consumed by the rules that reason about sampling operations.
We introduce a rule \rref{Laplace-shift} to reason about a pair of Laplacians,
with the same parameter $\err$ and means $m,m'$. By setting \(k=0\)
(and thus, $\abs{\lapmean - \lapmean'} \leq \scale$),
the rule lets us spend $\uptom{\scale \cdot \err}$ to ensure that both Laplacians
return the same result. 
If, instead, \(k > 0\) and \(\scale = 0\) (and thus, $\abs{\lapmean-\lapmean'} = k$),
the rule lets us conclude, without consuming privacy, that we will get
two samples at distance $k$. The rule combines the two reasoning principles into one:

\begin{equation*}
\infrule{Laplace-shift}
  {\abs{k + \lapmean - \lapmean'} \leq \scale }
  {\hoareref
    {\uptom{\scale \cdot \err}}
    {\Laplace \err\ \lapmean}{\Laplace \err\ \lapmean'}
    {z\ z'\,.\, z' = z+k}}
\end{equation*}
A consequence of \rref{Laplace-shift} (when $k=0$)
is that \(\Laplace\) is internally private, \ie, the \thelogic statement \(\hoaredppure{\err}{\Fun \vx . \Laplace\ \err\ \vx}\) is derivable.
This internalizes \DP of the Laplacian (\Cshref{thm:laplace-dp}).

Finally, we have the following novel rule for the Laplacian which can be used to
\emph{recover} some privacy credits. The idea is to partition the results into
two groups of outcomes depending on some threshold \(T\). If both Laplacians
sample a result above \(T\) (in fact, above \(T+1\) for the right-hand
side) the privacy credits are consumed. If, on the other hand, both results
remain below their respective thresholds, then the privacy credits can be
recovered.
This yields a useful reasoning principle for the Laplacian that is applied, \eg, in the verification of the \texttt{AboveThreshold} mechanism (\cref{sec:at}).
\begin{equation*}
  \infrule{Laplace-choice}
  {\abs{\lapmean-\lapmean'} \leq 1 \\ T \in \integer }
  {\hoareref
    {\uptom{2 \err}}
    {\Laplace \err\ \lapmean}{\Laplace \err\ \lapmean'}
    {z\ z'\,.\,
      \begin{aligned}
        &(T \leq z \,\land\, T+1 \leq z')
          \quad \lor \\[-0.6em]
        &(z < T \,\land\, z' < T+1 \,\sep\, \uptom{2 \err})
      \end{aligned}
    }}
\end{equation*}
Intuitively this rule is sound because it partitions the outcomes into disjoint events such that the total privacy budget for the two is bounded by the initial privacy budget.
Despite its conceptual simplicity, constructing a sound model that validates \rref{Laplace-choice} required substantial new insights in the form of a new composition theorem (see \cref{thm:choice-comp}).

\subsection{Internal Composition Laws}

Working in a program logic makes it simple to internalize the various composition laws for \DP,
as presented in \Cref{sec:preliminaries}.
First, we remark that the internal version of post-processing holds. This
is stated in our logic through the following lemma:

\begin{lemma}[Internal post-processing]
  \label{thm:internal-postp}
  Let \(\vf\) be internally \((\err,\aerr)\)-\dipr and assume \(\vg\) is ``safe to execute'' in the sense that
  $\All w. \hoareref{\TRUE}{\vg\, w}{\vg\, w}{\val\, \val' \ldotp \val = \val'}$ holds.
  Then \(\vg \circ \vf\) is internally \((\err,\aerr)\)-\dipr.%
\end{lemma}

This lemma can be proven entirely within the logic as an immediate consequence of the fact that \thelogic validates the \rref{Bind} rule.
Similarly to the definition of internal \DP, we can also recast the notion of sensitivity (\cshref{def:sens}) in \thelogic.
\begin{definition}
  A \thelang function \(\vf\) is \defemph{internally \(\scale\)-sensitive} (\(\hoaresens \scale \vf\)) if the following holds:
  \[
    \All x\, y : A . \hoareref{\TRUE}{\vf\, x}{\vf\, y}{\val\, \val' \ldotp \dr[B] \val {\val'} {\scale \cdot \df[A] x y}}
  \]
\end{definition}
As before, the definition is parametrized by two distances at types
$A$ and $B$, which can then be internalized as distances between values. The
internal equivalent of metric composition then follows.
\begin{lemma}\label{thm:imetric-comp}
  \(\All\, \vf\, \scale\, \err. \hoaresens \scale \vf \wand \hoaredppure{\err}{\Fun \vx . \Laplace\ {(\err/\scale)}\ (\vf\ \vx)}\) is derivable in \thelogic.
\end{lemma}

%% file: case-studies.tex
\section{Reusable Specs for Privacy Mechanisms}
\label{sec:case-studies}

Privacy \emph{mechanisms} are the primitive building blocks of \DP. Mechanisms sample the appropriate noise for a given data processing task. We give specifications for some widely used mechanisms and demonstrate that clients can be verified based on these abstract specs.

Our first example will be the Above Threshold (AT) mechanism, and we will see how to (1) prove that it satisfies an abstract specification capturing its privacy, (2) build the Sparse Vector Technique (SVT) from it, and (3) use AT to calculate the clipping bounds required to privately compute averages over a dataset. Both (2) and (3) use the same abstract specification of AT.

Besides the case studies presented hereafter, we also verified the privacy of the Report Noisy Max mechanism, which is of note because---as with AT---its privacy does not follow from composition laws but requires careful manipulation of the privacy budget and the use of budget- and state-dependent invariants. Details can be found in \appref{sec:report-noisy-max}.
We will present the most interesting ideas for each proof. Full proofs from the rules of \thelogic are available in our \rocq formalization.

\subsection{Sparse Vector Technique}
\label{sec:sparse-vector}

Suppose we have an incoming sequence of (1-sensitive) queries on a database and a fixed privacy budget. The Sparse Vector Technique (SVT) allows us to fix a threshold T and release, in a private manner, whether the result of each query exceeds T or not. The benefit of SVT is that one only has to spend privacy budget on the ``successful'' queries which do indeed exceed T and are released; the results of the queries that do not exceed the threshold can be computed (and discarded) without incurring a privacy cost. We can thus set in advance a maximum number $N$ of successful queries to be released, and keep answering incoming queries interactively until $N$ is reached.
The SVT is usually implemented in terms of the Above Threshold mechanism (AT), which finds a \emph{single} query above \(T\). SVT then simply runs \(N\) iterations of AT.
The privacy cost of SVT is \(N\) times the cost of finding one query above T.

The SVT is of interest for verification because several buggy privacy proofs have been published (see \citet{Lyu:Understanding:2017} for a survey). As the survey explains, the SVT is particularly interesting in the interactive setting; in the non-interactive setting, one can use the Exponential Mechanism instead and get more accurate results.
We explore two subtleties of SVT: how to perform the privacy analysis of AT, and how to build an interactive algorithm out of AT.
The privacy analysis of AT (and SVT) is challenging because it requires fine-grained reasoning about the privacy budget that \emph{cannot} be justified by the sequential composition law alone.

\subsubsection{Above Threshold}
\label{sec:at}

The Above Threshold mechanism can be used to evaluate queries on a database until one
query returns a result that exceeds a specified threshold~\(\vT\).
The implementation of \(\AT\) in \Cref{fig:at} initializes the noisy threshold \(\That\) and returns a function
to run queries interactively. This function receives a query, computes its result, adds additional noise to it
and checks whether the noisy result exceeds \(\That\).

\begin{figure}[thbp]
  \centering \small
  \renewcommand{\arraystretch}{1.05}
    \begin{tabular}[t]{>{$}l<{$} @{\;\;} >{$}l<{$} @{} >{$}l<{$} }
      \DumbLet\,\, \AT\ \verr\ \vT =
                   &\Let \That =\Laplace \ (\verr\, /\, 2) \ \vT in \\
                   &\DumbLet \vf\ \vq\ \vdb = \Let \vx = \vq  \ \vdb in \Let \vy = \Laplace \ (\verr\, /\, 4) \ \vx in \That \leq \vy \\
                   &\DumbIn \vf
  \end{tabular}
  \caption{The Above Threshold mechanism.}
  \label{fig:at}
\end{figure}

\begin{example}
  Suppose we want to privately compute the number of even numbers in a list
	and check whether it exceeds a threshold of 3. If we run AT with a high privacy budget
	(\eg, $\err=10$) there is very little noise added and we are very likely to disclose the true result and disclose private information. Concretely, with probability about 92\%,
    \\[1mm]
    {\small
      \renewcommand{\arraystretch}{1.05}
      \begin{tabular}[t]{>{$}l<{$} @{} >{$}l<{$} @{} >{$}l<{$} @{} >{$}l<{$} }
        &\AT\ 10\ 3\ (\listcount\ (\Fun x . x\ \langkw{mod}\ 2 = 0 ))\ & \quad [1,2,3,4,5] &\quad\to\quad \False \\
        &\AT\ 10\ 3\ (\listcount\ (\Fun x . x\ \langkw{mod}\ 2 = 0 ))\ &\quad [1,2,3,4,5,6] &\quad\to\quad \True
      \end{tabular}}\\

	As the value of \(\verr\) \emph{decreases}, we are more likely to observe $\True$ in the first query or $\False$ in the second, \ie, privacy improves while accuracy deteriorates.
\end{example}

Our specification~\eqref{eq:at-spec} captures the idea that \(\AT\) is \emph{interactive}: after initialization, the function \(\vf\) can be used to compare a query to the (noisy) threshold~\(\That\) until a result above~\(\That\) is found, but queries can be supplied and \emph{chosen} one by one after observing the result of previous queries.
\begin{align}\label{eq:at-spec}
  \hoareVref{\uptom{\verr}}
  {\AT\ \verr\ \vT}
  {\AT\ \verr\ \vT}
  {\vf\ \vf'\,.\,
  \begin{aligned}
    &\Exists AUTH . AUTH \sep \\[-0.6ex]
    &\quad \All\ \vdb\ \vdb'\ \vq . \hoareVref{AUTH \sep \adj{\vdb}{\vdb'} \sep \sens{1}{\vq}}{\vf\ \vq\ \vdb}{\vf'\ \vq\ \vdb'}{b\ b'\,.\ b = b' \sep \If \Negb b\ then AUTH}
  \end{aligned}
  }
\end{align}
Let us unpack the specification piece by piece. After initializing the mechanism with a privacy budget of \(\verr\), we obtain a pair of functions \(\vf\) and \(\vf'\), where both functions represent the same computation but with a priori different randomly sampled values of \(\That\), as well as an abstract ``authorization token'' \(AUTH\). The Hoare quadruple for \(\vf,\vf'\) in the postcondition indicates that so long as we have the \(AUTH\) token, running the functions on a 1-sensitive query \(\vq\) and adjacent datasets (1)~produces the same result for both \(\vq\ \vdb\) and \(\vq\ \vdb'\), \ie, the computation is private, and (2)~only consumes the \(AUTH\) token if \(\vq\ \vdb\) is above \(\That\). If the result of the comparison is \(\False\), \(AUTH\) can be recovered in the postcondition and we can continue to privately look for a query that exceeds~\(\That\). Note that the initial budget $\uptom{\verr}$ is only spent once at initialization, but we can still privately run as many queries as it takes to get a result above the threshold.

\paragraph{Proving Privacy} The proof of the specification~\eqref{eq:at-spec} proceeds as follows:
\begin{enumerate}
\item We split \(\uptom \verr\) into \(\uptom {\verr/2} \sep \uptom {\verr/2}\) and use the first half to pay for the \rref{Laplace-shift} rule with parameters \(\lapmean=\lapmean'=\vT,\scale = 1, k = 1\). We thus obtain related results \(\That, \That'\) for the left- and right-hand program such that \(\That' = \That+1\). Forcing the two noisy thresholds to be at distance 1 rather than equal is a standard ``trick'' in the analysis of Above Threshold.
\item We pick \(AUTH \eqdef \uptom{\verr/2}\) and use the remaining budget to provide the initial \(AUTH\) token.
\item We now have to show the specification for
  {
    \setlength{\abovedisplayskip}{5pt}
    \setlength{\belowdisplayskip}{5pt}
    \begin{align*}
    \vf\phantom{'}\ \vq\ \vdb\phantom{'} &\;\eqdef{}\;
          \Let \vx\phantom{'} = \vq\ \vdb\phantom{'} in
          \Let \vy\phantom{'} = \Laplace \ (\verr\ /\ 4) \; \vx\phantom{'} in
          \That\phantom{'} \leq \vy\phantom{'} \\
    \vf'\ \vq\ \vdb' &\;\eqdef{}\;
          \Let \vx' = \vq\ \vdb' in
          \Let \vy' = \Laplace \ (\verr\ /\ 4) \; \vx' in
          \That' \leq \vy'
  \end{align*}}
\item By sensitivity of \(\vq\) and adjacency of the databases, \(\vx\) and \(\vx'\) are at distance at most 1.
\item By the definition of \(AUTH\), the precondition of the refinement confers us a privacy budget of \(\uptom{\err/2}\). We use this budget to apply the \rref{Laplace-choice} rule and partition the outcomes \((\vy, \vy')\) of the remaining Laplace sampling into two mutually exclusive cases:
  \begin{itemize}
  \item \( \That \leq \vy \) and  \( \That' \leq \vy' \). In this case, both of the comparisons return \(\True\).
  \item \(\vy < \That\) and  \(\vy' < \That'\). In this case, both comparisons return \(\False\), the rule does not consume the privacy budget and we can return \(AUTH\).
  \end{itemize}
  Either way, \(\vf\) and \(\vf'\) return the same result \(b\), and if \(b=\False\) then \(AUTH\) is returned too. \qed
\end{enumerate}
To see that the specification~\eqref{eq:at-spec} is indeed useful, we will now use it to verify the privacy of the interactive sparse vector technique.

\subsubsection{An interactive Sparse Vector Technique}
\label{sec:isvt}

We prove privacy of the interactive SVT directly from the abstract specification of the AT mechanism \eqref{eq:at-spec}. SVT orchestrates repeated invocations of \(\AT\) in order to identify the first \(\vN\) of the queries that exceed the threshold \(\vT\). Unlike a purely batch-style algorithm, the SVT exposes a streaming interface that allows queries to be submitted interactively, \ie, depending on the results of earlier queries. This makes the verification of privacy significantly more challenging, as the set of queries cannot be fixed in advance but may depend on previously released (noisy) information.
In \thelogic however, the proof that \(\SVT\) is %
\dipr is a  relatively straightforward consequence of the privacy of \(\AT\).

The implementation in \Cref{fig:svt} works as follows. It maintains two pieces of mutable state: a reference \(\vAT\) to the current Above Threshold instance and a counter that tracks how many additional \(\True{}\) results (\ie, queries that exceed the threshold) may still be released. Each invocation of the returned function \(\vf\) runs the current \(\AT\) instance on a new query \(\vq\) and database \(\vdb\), producing a boolean result \(\vb\). If \(\vb\) is \(\True\) and the counter has not yet reached zero, the mechanism consumes \(\verr\) privacy credits and reinitializes \(\vAT\) with a fresh Above Threshold instance. Otherwise, the counter and function reference remain unchanged. The caller may then use the result \(\vb\) to decide which query to issue next.

\begin{figure}[htbp]
  \centering \small
  \renewcommand{\arraystretch}{1.05}
  \begin{minipage}[t]{0.525\textwidth}
    \begin{tabular}[t]{>{$}l<{$} @{} >{$}l<{$}}
      \DumbLet\,\, &\SVT\ \verr\ \vT\ \vN =\\
      &
        \begin{tabular}[t]{>{$}l<{$} @{} >{$}l<{$} @{} >{$}l<{$} }
          \Let \vAT = \Alloc{} (\AT\ \verr\ \vT) in \\
          \Let \vcounter = \Alloc{} (\vN - 1) in \\
          \DumbLet \vf\ \vq\ \vdb =
          \begin{tabular}[t]{@{}>{$}l<{$} @{} >{$}l<{$} @{} >{$}l<{$} }
            \Let \vb = (\deref \vAT) \; \vq \ \vdb in \\
            \If \deref \vcounter > 0 \,\And\, \vb  then \\
            \tab (\;\vcounter \gets (\deref \vcounter - 1) \Seq \\
            \tab \phantom{(}\; \vAT \gets \AT\ \verr\ \vT\;) \Seq \phantom{)} \\
            \vb
          \end{tabular}
          \\
          \DumbIn \vf
        \end{tabular}
    \end{tabular}
  \end{minipage}
  \hfill
  \begin{minipage}[t]{0.445\textwidth}
    \begin{tabular}[t]{>{$}l<{$} @{} >{$}l<{$} @{} >{$}l<{$} }
      \DumbLet\,\, &\SVTstream\ \verr\ \vT\ \vN\ \vqstream\ \vdb = \\
                   &\Let \vf = \SVT\ \verr\ \vT\ \vN in \\
                   &\DumbLet \Rec {\langv{iter}} {\vi\ \vbs} = \\
                   &\tab \If \vi = \vN then \listrev\ \vbs \\
                   &\tab \Else \Let \vq = \vqstream\ \vbs in \\
                   &\tab \phantom{\Else} \Let \vb = \vf\ \vq\ \vdb in \\
                   &\tab \phantom{\Else} \langv{iter}\ (\If \vb then (\vi+1) \Else \vi)\ (\vb :: \vbs)
      \\
                   &\langkw{in}\spac \langv{iter}\ 0\ []
    \end{tabular}
  \end{minipage}
    \label{fig:svt}
    \caption{The Sparse Vector Technique and a streaming client.}
\end{figure}

The specification~\eqref{eq:svt-spec} formalizes the intuition that the interactive SVT behaves like a private state machine that can be queried multiple times while consuming a fixed privacy budget.
\begin{align}\label{eq:svt-spec}
  \hoareVref{\uptom{\vN \cdot \verr}}
  {\SVT\ \verr\ \vT\ \vN}
  {\SVT\ \verr\ \vT\ \vN}
  {\vf\ \vf'\,.\,
  \begin{aligned}
    &\Exists\ \iSVT . \iSVT(\vN) \sep \\[-0.6ex]
    &\quad \All\ \vdb\ \vdb'\ \vq\ n. \hoareVref{\iSVT(n+1) \sep \adj{\vdb}{\vdb'} \sep \sens{1}{\vq}}{\vf\ \vq\ \vdb}{\vf'\ \vq\ \vdb'}{b\ b'\,.\ b = b' \sep \iSVT (\If b then n \Else n+1)}
  \end{aligned}
  }
\end{align}
Initially, the mechanism owns a total privacy resource of  \(\uptom{\vN\cdot\verr}\), corresponding to~\(\vN\) possible above-\(\vT\) releases. The abstract token \(\iSVT(K)\) in the postcondition tracks the remaining number $K$ of \(\True\) results we can release.
The returned function \(\vf\) satisfies the nested quadruple: if we own at least $\iSVT(1)$, we can run a 1-sensitive query on adjacent databases and ensure we get the same result; if the result is true (i.e., the query exceeds the threshold), we decrease $\iSVT$ by 1, otherwise we get back our initial token.
This specification thus captures both the privacy accounting and the interactive behavior of SVT: the mechanism remains private for any adaptively chosen sequence of 1-sensitive queries until the allotted number \(\vN\) of positive releases has been exhausted.

Notably, the proof of this specification treats AT abstractly and relies only on its specification~\eqref{eq:at-spec}, not its implementation,
which gives us a more modular analysis. The proof relies on keeping a simple invariant over the counter, the AT reference and the remaining privacy
budget.

To summarize, this implementation and our verification of it has several noteworthy features:
\begin{itemize}
  \item the SVT is a client of AT via an abstract specification,
  \item its interactive interface is specified through nested Hoare quadruples, and
  \item it requires storing a private higher-order function in a reference.%
\end{itemize}

Next, we demonstrate that our specifications are expressive enough for reuse
by different clients.

\subsubsection{SVT Client: Streams of Queries}
\label{sec:svt-stream}

The standard textbook account of SVT \cite{Dwork:Algorithmic:2014} presents it as an algorithm that takes in a \emph{stream} of queries \(\vqstream\) and produces a list of booleans \(bs\) where \(b_i\) indicates whether the \(i\)-th query was above the noisy threshold. The stream is represented as a (possibly stateful) function that produces a new query on each invocation, and interactivity is modeled by the fact that each time a new query is requested, \(\vqstream\) gets access to the booleans \(bs\) resulting from the preceding queries.

We can directly prove that the implementation \(\SVTstream\) (\Cshref{fig:svt}) is private by applying the generic specification for SVT. A user of \(\SVTstream\) only has to satisfy the textbook assumption that all of the queries are indeed 1-sensitive. The privacy reasoning is encapsulated in~\eqref{eq:svt-spec}.

\subsubsection{Above Threshold Client: auto\_avg}
\label{sec:auto-avg}

As a last application of the Above Threshold mechanism, we analyze the \(\vautoavg\) client that privately computes the average of a dataset.
This example is taken from the online textbook \cite[Chapter~10]{Near:Programming:2021}.

The fact that both \(\SVT\) and \(\vautoavg\) can both be verified against the same abstract interface for \(\AT\) is good evidence that our specifications are indeed reusable and can be used to verify libraries without having to worry about implementation details.

\begin{figure}[thbp]
  \centering
  \renewcommand{\arraystretch}{1.05}\small
  \begin{minipage}[t]{0.534\linewidth}
    \begin{tabular}[t]{>{$}l<{$} @{} >{$}l<{$} @{} >{$}l<{$} }
      &\DumbLet \vautoavg\ \vbnds\ \verr\ \vdb = \\
      &\tab \Let \vbound = \vclipbound\ \vbnds\ \verr\ \vdb in \\
      &\tab \Let \vsum = \vclipsum\ \vbound\ \vdb in \\
      &\tab \Let \vsumnoisy  = \Laplace\ (\verr\ /\ \vbound)\ \vsum in \\
      &\tab \Let \vcountnoisy = \Laplace\ \verr\ (\listlen\ \vdb) in \\
      &\tab \vsumnoisy\ /\ \vcountnoisy \\[1.5mm]
      &\DumbLet \vclipbound\ \vbnds\ \verr\ \vdb = \\
      &\tab \Let \vqs = \listmap\ (\Fun \vb . (\vb, \vcreatequery\ \vb))\ \vbnds in \\
      &\tab \Let (\vbound, \_) = \vatlist\ \verr\ 0\ \vdb\ \vqs in \\
      &\tab \vbound
    \end{tabular}
  \end{minipage}
  \hfill
  \begin{minipage}[t]{0.46\linewidth}
    \begin{tabular}[t]{>{$}l<{$} @{} >{$}l<{$} @{} >{$}l<{$} }
      &\DumbLet \vcreatequery\ \vb\ \vdb = \\
      &\tab (\vclipsum\ \vb\ \vdb) - (\vclipsum\ (\vb + 1)\ \vdb) \\[1.5mm]
      &\DumbLet \vclipsum\ \vbound\ \vdb = \\
      &\tab \listsum\ (\listclip\ \vbound\ \vdb) \\[1.5mm]
      &\DumbLet \vatlist\ \verr\ \vT\ \vdb\ \vqs = \\
      &\tab \Let \vAT = \AT\ \verr\ \vT in \\
      &\tab \listfind\ (\Fun (\vbound, \vq). \vAT\ \vq\ \vdb)\ \vqs
    \end{tabular}
  \end{minipage}
  \caption{Privately computing the average of a list of data.}
  \label{fig:autoavg}
\end{figure}

To privately compute the average of a dataset it is not enough to first compute the average and then add \(\verr\) Laplacian noise for a fixed \(\verr\), as this may leak information about the size of the dataset.
The noise has to be calibrated to the largest element---but that value in itself is private information!

The solution adopted in the implementation of \(\vautoavg\) in \Cref{fig:autoavg} is to ``clip'' the elements of the database to lie in a bounded range \([0, B]\). If two adjacent databases are clipped to the same bound, their sum can differ by at most \(B\).
In other words, we can prove that \(\vclipsum\ B\) is internally \(B\)-sensitive.
We can apply internal metric composition (\Cshref{thm:imetric-comp}) to show that computing \(\vsumnoisy\) by adding to the clipped sum Laplacian noise with scale \((\verr/B)\) is \(\verr\)-private.
Therefore, \(\vautoavg\) achieves \((3{\cdot}\verr)\)-\DP, where the budget is divided equally between the call to \(\vclipbound\) and the two calls to \(\Laplace\):
\begin{equation*}
  \hoareref{\uptom{3 \cdot \verr} \sep \adj{\vdb}{\vdb'}}{\vautoavg\ \vbnds\ \verr\ \vdb}{\vautoavg\ \vbnds\ \verr\ \vdb'}{x\ x' \ldotp x = x'}
\end{equation*}

The utility of \(\vautoavg\) stems from carefully choosing \(B\).
Given a list of candidate bounds \(\vbnds\) we can do this privately via the AT mechanism.
The function \(\vclipbound\) to finds the first value \(\vb\) in \(\vbnds\) such that the sum of elements in \(\vdb\) stops increasing if the clipping bound is relaxed from \(\vb\) to \(\vb+1\). Testing this for all values in \(\vbnds\) via \(\vcreatequery\) is 1-sensitive. Therefore we can directly apply the specification~\eqref{eq:at-spec} for \(\AT\) to derive that \(\vclipbound\) is \(\verr\)-\dipr.

\subsection{Privacy Filters}
\label{sec:privacy-filters}

A common implementation technique for \DP in general-purpose programming languages is to explicitly track the remaining privacy budget as a program variable. At the beginning of a data analysis, this variable is initialized to the global privacy budget \(\err\), and it must remain non-negative throughout the program execution. So long as care is taken to decrement the budget every time (noisy) data is released, the whole data analysis is \(\err\)-private.
To ensure that these rules are respected, the management of the privacy budget is commonly encapsulated in a \emph{privacy filter}, a higher-order function that runs a computation only if there is sufficient budget for it.
This programming pattern provides a separation of concerns: if the privacy analysis of the individual computations is correct, and the filter is correctly implemented,
then the entire computation is private. This allows us to verify the different components (privacy filter and mechanisms) modularly.

\begin{figure}[thbp]
  \centering \small
  \renewcommand{\arraystretch}{1.1}
  \begin{minipage}[t]{0.31\linewidth}\hspace{-1ex}
    \begin{tabular}[t]{>{$}l<{$} @{} >{$}l<{$} @{} >{$}l<{$} }
      \DumbLet\,\, &\PF\ \vbudget\ = \\
                   &\Let \vbudrem = \Alloc{} \vbudget in \\
                   &\DumbLet \vtryrun\ \vcost\ \vf = \\
                   &\tab\begin{tabular}[t]{>{$}l<{$} @{} >{$}l<{$} @{} >{$}l<{$} }
                     &\If \deref \vbudrem < \vcost then\\
                     &\tab \None \\
                     &\Elsenospac \\
                     &\tab \vbudrem \gets \deref \vbudrem - \vcost \Seq \\
                     &\tab \Some\ (\vf\ \TT)
                   \end{tabular} \\
                   &\langkw{in}\spac \vtryrun
    \end{tabular}
  \end{minipage}
  \hfill
  \begin{minipage}[t]{0.64\linewidth}\hspace{-1ex}
    \begin{tabular}[t]{>{$}l<{$} @{} >{$}l<{$} @{} >{$}l<{$} }
      \DumbLet\,\, &\ACount\ \verrcoarse\ \verrprecise\ \vT\ \vbudget\ \vpreds\ \vdb = \\
                   &\Let \vtryrun = \PF\ \vbudget in \\
                   &\listmap\ (\Fun \vpred .
                     \begin{tabular}[t]{@{}>{$}l<{$} @{} >{$}l<{$} @{} >{$}l<{$} }
                       &\Let \vcountexact = \listcount\ \vpred\ \vdb in  \\
                       &\Let \vg\ \_ = \Laplace\ \verrprecise\ \vcountexact in  \\
                       &\DumbLet \vf\ \_\, =
                         \begin{tabular}[t]{@{}>{$}l<{$} @{} >{$}l<{$} @{} >{$}l<{$} }
                           & \Let \vcountcoarse = \Laplace\ \verrcoarse\ \vcountexact in  \\
                           & \DumbLet \vcountprecise = \langkw{if}\spac \vT <  \vcountcoarse  \\
                           &\tab \langkw{then}\spac\, \vtryrun\ \verrprecise\ \vg \\
                           &\tab \Elsenospac\,\, \None \In \\
                           & (\vcountcoarse,\, \vcountprecise)
                         \end{tabular} \\
                     \end{tabular} \\
                   &\tab\langkw{in}\spac \vtryrun\ \verrcoarse\ \vf) \\
                   &\tab\vpreds
    \end{tabular}
  \end{minipage}
  \caption{Implementations of a Privacy Filter and Adaptive Count.}
  \label{fig:filter}
\end{figure}

The implementation \(\PF\) in \Cref{fig:filter} works as follows. Upon initialization it allocates a reference that tracks the remaining privacy budget and returns a closure \(\vtryrun\) that can be used to run private computations in an interactive manner.
If a client tries to run a computation with cost exceeding the remaining budget, \(\vtryrun\) does not run the computation, otherwise
it decreases the budget by the cost, runs the computation, and returns the result.

For the sake of simplicity we present a privacy filter that only tracks the \(\err\)-budget, but the method directly generalizes to \((\err,\aerr)\)-privacy filters.

\subsubsection{Proving Privacy of Privacy Filters}
\label{sec:provove-filter}

The high-level intuition for the privacy filter is that it should never exceed the budget that was initially set up so long as any client that calls \(\vtryrun\ \vcost\ \vf\) ensures that \(\vf\) is indeed \(\vcost\)-\dipr. This intuition is captured by the following specification.
\begin{align}
  \label{eq:filter-spec}
  &\All\, \vbudget\, .
  &&\!\!\! \hoareVref{\uptom{\vbudget}}{\PF\ \vbudget}{\PF\ \vbudget}
    { \vtryrun\ \vtryrun'\,.\, \Exists\ \iPF . \iPF \sep \vtryrun{}\textit{-spec} }
  \\
  \intertext{where \(\vtryrun\)\textit{-spec} is defined as}
  \label{eq:tryrun-spec}
  &\All\, \vcost\ \vf\ \vf'\ \iIf \, . %
  &&\!\!\! \raisebox{8.4pt}{\(\displaystyle \hoareVref{
    \begin{aligned}
      &\iPF \sep \iIf \\
      &\; \sep\hoareref{\uptom {\vcost} \sep \iPF \sep \iIf}{\vf\ \TT}{\vf'\ \TT}{\val\ \val'\,.\ \val = \val' \sep  \iPF \sep \iIf}
    \end{aligned}
    }
    {\vtryrun\ \vcost\ \vf}
    {\vtryrun'\ \vcost\ \vf'}
    {b\ b'\,.\ b = b' \sep \iPF \sep \iIf}\)}
\end{align}
The existentially quantified \(\iPF\) token in the postcondition of \eqref{eq:filter-spec} represents a client's ability to execute computations privately via \(\vtryrun\).
The specification \eqref{eq:tryrun-spec} defines the behavior of \(\vtryrun\ \vcost\ \vf\). Assuming that for a privacy cost of \(\vcost\) the functions \(\vf\) and \(\vf'\) produce equal results and maintain an invariant \(\iIf\), calling them through \(\vtryrun\) (and \(\vtryrun'\) respectively) satisfies the same invariant regardless of whether there actually is enough budget left to execute \(\vf\). Since \(\vf\) may itself contain calls to \(\vtryrun\), the specification for \((\vf, \vf')\) has access to the \(\iPF\) token. We give a simple application of this expressivity in the form of nested calls to the privacy filter in \Cref{sec:adaptive-count}.

The proof of the specification in \thelogic is straightforward. We define the token \(\iPF\) as
\begin{equation*}
  \Exists \err . \uptom{\err} \;\sep\; \vbudrem \pointsto \err \; \sep \; \vsbudrem \spointsto \err \;.
\end{equation*}
The link between the logical resource representing ownership of the error budget and the program state tracking the remaining budget allows us to conclude that the call to \(\Some\ (\vf\ \TT)\) in the definition of \(\vtryrun\) is only executed when sufficient privacy budget remains to satisfy the precondition of \((\vf,\vf')\) in \eqref{eq:tryrun-spec}, and hence the invariant \(\iIf\) is satisfied. In case \(\vtryrun\) decides that the budget is insufficient for \(\vf\), the invariant is trivially preserved.

The power of this specification for \(\PF\) lies in the fact that a client does not have to perform \emph{any} privacy accounting or reasoning whatsoever for \(\vtryrun\ \vcost\ \vf\)! We can conveniently combine library functions to build a private \(\vf\) and run it without having to worry whether we still have enough budget: the filter  ensures that the initial budget is never exceeded.

\subsubsection{Client: Adaptive Counting}
\label{sec:adaptive-count}

Another advantage of implementing \DP through privacy filters is that it allows the data analyst to decide \emph{dynamically} where the privacy budget should be spent, \ie, the way the budget is spent can adapt to the results of prior analyses.
This is especially useful in exploratory data analysis when it is unclear, a priori, what values the dataset ranges over.

The example \(\ACount\) in \Cref{fig:filter} employs a form of adaptivity to privately count the number of elements of \(\vdb\) that satisfy each of the tests in the list of \(\vpreds\).
First, a cheaper but less precise count is performed, consuming \(\verrcoarse\) privacy credits. Only if this yields a promising result that exceeds a threshold \(\vT\), a more precise analysis is performed for an additional larger budget of \(\verrprecise\). The result of \(\ACount\) is thus a numeric estimate for each predicate, with a more precise value for a few ``important'' candidates.

A conservative privacy analysis would have to assume a worst-case cost of \(\length(\vpreds) \cdot ( \verrcoarse + \verrprecise)\) even if many of the coarse counts may in practice not exceed the threshold.
By employing a privacy filter, we can instead fix a budget \(\vbudget\) that we want to allocate to this analysis task and try to run the analysis so long as the filter has enough budget left. If only a few of the data entries exceed \(\vT\), this allows to count many more predicates than the conservative analysis without a privacy filter.

In \thelogic, we can prove that \(\ACount\) is \(\vbudget\)-\dipr from the specification \eqref{eq:filter-spec} because \(\vf\) and \(\vg\) meet the precondition of \(\vtryrun\), as they  consume a budget of \(\verrcoarse\) and \(\verrprecise\) respectively.

\subsection{Caching Techniques for \DP}
\label{sec:cache}

Interactive analysis with data-dependent queries is common in real-world workloads for \DP.
This poses a challenge for \DP frameworks because it makes it impossible to statically avoid repeated evaluation of certain queries, say, by refactoring code that requires the same result, and hence repeated queries inflate the privacy cost unnecessarily.
This problem can be solved with a query cache that memoizes the results of a query on first execution and reuses this result upon repetition.
One would hope that reusing a noisy result in a repeated query should be ``for free'' and consume no privacy budget.
However, the privacy analysis of such a memoization method is subtle, because the privacy cost of a query depends on the history of queries, which is highly non-local information.

\begin{figure}[thbp]
  \centering \small
  \renewcommand{\arraystretch}{1.05}
  \begin{minipage}[t]{0.46\linewidth}%
    \begin{tabular}[t]{>{$}l<{$} @{} >{$}l<{$} @{} >{$}l<{$} }
      \DumbLet\,\, &\vmkcache\ \vaddnoise\ \vdb\ = \\
                   &\Let \vcache = \mapinit\ \TT in \\
                   &\DumbLet \vruncache\ \vq = \\
                   &\tab
                     {\MatchMLNoEnd \mapget\ \vcache\ \vq  with
                     | \Some\, \vx => \vx
                     | \None =>
                     \begin{tabular}[t]{>{$}l<{$}}
                       \Let \vx = \vaddnoise\ \vq\ \vdb in \\
                       \mapset\ \vcache\ \vq\ \vx \Seq \\
                       \vx
                     \end{tabular}
                     end {}}
                   \\
                   &\langkw{in}\spac \vruncache
    \end{tabular}
  \end{minipage}
  \hfill
  \begin{minipage}[t]{0.53\linewidth}%
    \begin{tabular}[t]{>{$}l<{$} @{} >{$}l<{$} @{} >{$}l<{$} }
      \DumbLet\,\, &\vmapcache\ \vaddnoise\ \vqs\ \vdb\ = \\
                   &\DumbLet \vruncache = \\
                   &\tab \vmkcache\ \vaddnoise\ \vdb \, \In \\
                   &\listmap\ \vruncache\ \vqs
    \end{tabular}
  \end{minipage}
  \caption{Implementations of a cache and a client.}
  \label{fig:cache}
\end{figure}

In this section, we reason about the privacy of the caching method introduced in
\Cref{fig:py-cache}, which we implement in \thelang through the algorithm shown in
\Cref{fig:cache}. Our formalization crucially relies on our logic being able to support
higher-order functions, local state, and a resource-based representation of the
privacy budget.

\subsubsection{Cache spec: repeated queries are free}
\label{sec:cache-spec}
Upon initialization, \(\vmkcache\) allocates a mutable map \(\vcache\) and returns a closure \(\vruncache\) that stores and looks up noisy query results in \(\vcache\).
This is reflected in \eqref{eq:cache-spec} as the existentially quantified \(\iCache(\cachemap)\) resource. Initially, the map \(\cachemap\) is empty, but it can be updated via \(\vruncache\) as we will see next.
\begin{align}
  \label{eq:cache-spec}
  &
    \quad
    \hoareVref{ \adj{\vdb}{\vdb'}}
    {\vmkcache\ \vaddnoise\ \vdb}
    {\vmkcache\ \vaddnoise\ \vdb'}
    {\vruncache\; \vruncache'.\spac \Exists \iCache . \iCache(\mapempty)
    \sep \textit{spec-cached}
    \sep \textit{spec-fresh} }
  \intertext{where \textit{spec-fresh} is defined as}
  \label{eq:runcache-fresh}
    \All\, \cachemap\ \vq\, .
  &
    \quad
    \hoareVref{
    \vq \not\in \dom(\cachemap)
        \sep \uptom {\err} \sep \upto {\aerr}
        \sep \hoaredp \err \aerr {\vaddnoise\ \vq}
    \sep \iCache(\cachemap)
    }
    {\vruncache\ \vq}
    {\vruncache'\ \vq}
    {\val\ \val'\,.\ \val = \val' \sep \iCache(\cachemap[\vq \mapsto \val])}
  \intertext{and \textit{spec-cached} is defined as}
  \label{eq:runcache-cached}
    \All\, \cachemap\ \vq\, .
  &
    \quad
  \hoareVref{
    \vq \in \dom(\cachemap) \sep \iCache(\cachemap)
    }
    {\vruncache\ \vq}
    {\vruncache'\ \vq}
    {\val\ \val'\,.\ \val = \val' \sep \iCache(\cachemap) \sep \cachemap[\vq] = \val}
\end{align}
The specification \eqref{eq:runcache-fresh} describes the behavior of \((\vruncache\, \vq)\) on a query that has not been memoized yet. It requires that \(\vaddnoise\) should indeed run \(\vq\) under \((\err,\aerr)\)-\(\hoaredpname\) and assumes ownership of enough privacy credits to pay for this execution. Furthermore, it requires ownership of \(\iCache(\cachemap)\) for the current internal state of the cache. In the postcondition, we recover \(\iCache\) where \(\cachemap\) is updated with the result of the noisy query.

The intuition that repeated queries should be free is formalized in~\eqref{eq:runcache-cached}: if \(\vq\) is in the cache then no privacy credits are consumed for executing it under \(\vruncache\).

\subsubsection{A cache client}
\label{sec:xc-offline}
A simple application is \(\vmapcache\) (\cref{fig:cache}), which employs the cache to run \(\vaddnoise\) on a list of queries~\(\vqs\).
We can prove that the privacy cost of \(\vmapcache\) is \((k \err, k \aerr)\) where \((\err,\aerr)\) is the privacy cost of the \(\vaddnoise\) mechanism and \(k = \card{\set{\vq \in \vqs}}\) is the number of unique queries in \(\vqs\).
The proof of \eqref{eq:cache-client} follows directly from the abstract specification \eqref{eq:cache-spec}.
\begin{align}
  \label{eq:cache-client}
  (\All \vq \in \vqs . \hoaredp{\err}{\aerr}{\vaddnoise\ \vq}) \wand
  \hoaredp{k\err}{k\aerr}{\vmapcache\ \vaddnoise\ \vqs}
\end{align}
Without caching, the privacy cost would have to be multiplied by \(\textlog{List.}\length(\vqs)\) instead of~\(k\).

%% file: model.tex
\section{Soundness: A Model of \thelogic}
\label{sec:model}

In this section we give an overview of the model behind \thelogic{} and its adequacy theorem.
A detailed account can be found in \appref{sec:app-model}.

Our program logic is based around the notion of $(\varepsilon,\delta)$-approximate coupling \cite{advanced_for_diff_privacy, Sato:Approximate:2016}:

\begin{definition}

Let $A, B$ be countable types, and $\Phi \subseteq A \times B$ a relation. Given
two real-valued parameters $0 \leq \varepsilon, \delta$, we say that there is
an $(\varepsilon, \delta)$-approximate $\Phi-$coupling between distributions
$\mu_1 \colon \Distr{A}, \mu_2 \colon \Distr{B}$ if, for any real-valued random
variables $f : A \to [0,1], g : B \to [0,1]$ such that $\forall (a,b)\in\Phi, f(a) \leq g(b)$,
the following holds : $\expect[\mu_1]{f} \leq \exp(\varepsilon) \cdot \expect[\mu_2]{g} + \delta$.
We denote the existence of such a coupling by $\dpcoupl{\mu_1}{\mu_2}{\varepsilon}{\delta}{\Phi}$.

\end{definition}

The model is similar in spirit to that of Approxis~\cite{approxis}, which can
be seen as based on a notion of $(0,\delta)$-approximate coupling. Crucially,
we prove a novel \emph{choice composition} theorem, that is key to some of our
proof rules, in particular \rref{Laplace-choice}. This is heavily inspired by
choice couplings \cite{Albarghouthi:Synthesizing:2017}, but it really shines in our
setting, since we have a resourceful treatment of the privacy budget.

\begin{theorem}\label{thm:choice-comp}
  Let $\mu_1 \colon \Distr{A}, \mu_2 \colon \Distr{B}, f \colon A \to \Distr{A'}, g \colon B \to \Distr{B'}$.
  Assume we have a predicate $\Xi \subseteq A$, and $\Phi_1, \Phi_2 \subseteq A\times B, \Psi \subseteq A'\times B'$,
  with $\Phi_1, \Phi_2$ disjoint in the sense that, $\forall a, a', b. a \in \Xi \wedge a' \not\in \Xi \Rightarrow ((a,b)\not\in \Phi_1 \vee (a',b)\not\in \Phi_2)$.
  Assume further that:
  \begin{enumerate}[label=(\roman*)]
  \item\label{it:cc-Xi} $\dpcoupl{\mu_1}{\mu_2}{\varepsilon_1}{\delta_1}{\Phi_1}$
  \item\label{it:cc-notXi} $\dpcoupl{\mu_1}{\mu_2}{\varepsilon_2}{\delta_2}{\Phi_2}$
  \item For all $a,b$ such that $a \in \Xi$ and $(a,b)\in\Phi_1$,\; $\dpcoupl{f\,a}{g\, b}{\varepsilon_1'}{\delta_1'}{\Psi}$.
  \item For all $a,b$ such that $a \not\in \Xi$ and $(a,b)\in\Phi_2$,\; $\dpcoupl{f\,a}{g\, b}{\varepsilon_2'}{\delta_2'}{\Psi}$.
  \end{enumerate}

	Then, $\dpcoupl{(\mu_1 \mbindi f)}{(\mu_2 \mbindi g)}{\varepsilon}{\delta}{\Psi}$, where
  $\varepsilon \triangleq {\sf max}(\varepsilon_1 + \varepsilon_1', \varepsilon_2 + \varepsilon_2')$
  and
  $\delta \triangleq \delta_1 + \delta_2 + {\sf max}(\delta_1', \delta_2')$

\end{theorem}

This statement might appear cryptic at first. The idea is that, when proving a
coupling for the composed computations $\mu_1 \mbindi f, \mu _2\mbindi g$, one
can choose between two different couplings for $\mu_1,\mu_2$, each with a
different cost $\varepsilon_1$ or $\varepsilon_2$, and then also spend a
different privacy cost $\varepsilon_1'$ or $\varepsilon_2'$ for the
continuations depending on whether the value sampled from $\mu$ falls inside or
outside $\Xi$ (note however, that the additive cost $\delta_1 + \delta_2$ for the
first step must be paid in any case).

For instance, the proof of \rref{Laplace-choice} for parameters \(\err, T, m, m'\) essentially uses \cref{thm:choice-comp} with \(\Xi(z) \eqdef (T \leq z)\), \(\varepsilon_1 \eqdef 2\varepsilon\), \(\varepsilon_1' \eqdef 0\), \(\varepsilon_2 \eqdef 0\), and \(\varepsilon_2' \eqdef 2\varepsilon\). The coupling \ref{it:cc-Xi} comes from \rref{Laplace-shift} with \(k = 1\), which (together with \(z \in \Xi\)) ensures the condition \(\Phi_1 \eqdef (T \leq z \land T+1 \leq z')\) in the postcondition of \rref{Laplace-choice} for a privacy cost of \(\varepsilon_1 = 2\varepsilon\) since \(\abs{k+m-m'} \leq 2\). The coupling \ref{it:cc-notXi} comes from \rref{Laplace-shift} with \(k \eqdef m' - m\), which (together with \(z \not\in \Xi\)) ensures the condition \(\Phi_2 \eqdef (z < T \land z' < T+1)\) in the postcondition of \rref{Laplace-choice} for a privacy cost of \(\err_2 = 0\). Since the coupling for \ref{it:cc-notXi} does not consume any privacy credits, we recover \(\err_2' = 2\err\) in the second branch of the postcondition of \rref{Laplace-choice}.

In the model of \thelogic{}, this
composition is possible at every execution step, and realized through the way
separation logic resources tracking the budgets are threaded through the execution.

There is a tight connection between
couplings and \DP: a function \(f : \DB \to \Distr X\) is \((\varepsilon,\delta)\)-\dipr
iff for any adjacent inputs $b,b'\colon \DB$, we have $\dpcoupl{f~b}{f~b'}{\varepsilon}{\delta}{(=)}$.
With this connection in mind, we can now state the adequacy theorem of \thelogic{}.

\begin{theorem}
  Let $f,f'$ be two \thelang functions and $\Phi, \Psi \colon \Val \times \Val \to \Prop$.
  If
  $$\hoareref{\Phi(w,w') \ast \upto{\delta} \ast \uptom{\varepsilon}}{f~w}{f'~w'} {\Psi}$$
  holds in \thelogic{} then, for any initial states $\sigma,\sigma'$, we have $\dpcoupl{\exec(f~w,\sigma)}{\exec(f'~w',\sigma')}{\varepsilon}{\delta}{\Psi}$.
\end{theorem}

By instantiating $\Phi$ with adjacency and $\Psi$ with equality, we recover \cref{thm:soundness-dp} as a corollary.

Internally, the Hoare quadruples are defined in terms of a primitive, unary
notion of weakest precondition (WP), where the right-hand side program is
represented as a separation logic resource~\cite{reloc, clutch, approxis}.
Validity of the WP is defined by guarded induction on the program execution,
establishing an approximate coupling at each step, and finally composing all
the couplings into a coupling for the full execution. Each program logic rule
is then proven sound \wrt the definition of the WP. In
particular, all standard separation logic rules for the deterministic fragment
of \thelang can be re-established. We refer the reader to the supplementary
material for more details.

%% file: related-work.tex
\section{Related Work}
\label{sec:related-work}

  \begin{figure}
    \centering
    \footnotesize
    \setlength{\tabcolsep}{4pt}
    {%
      \newcommand\yes{\CIRCLE}
      \newcommand\no{\Circle}
      \newcommand\stglobal{\(\LEFTcircle\)%
      }
      \newcommand\stlocal{\(\CIRCLE\)%
      }
      \newcommand\stmonad{\(\no\)%
      }
      \newcommand\stfo{\(\triangle\)}
      \newcommand\stho{\(\blacksquare\)}
      \newcommand\stnone{\(\Circle\)}
      \newcommand\old{\LEFTcircle}
      \newcommand{\iadv}{\LEFTcircle}
      \newcommand\appdp{\((\merr, \aerr)\)}
      \newcommand\recwhile{}
      \hspace*{-1cm}
      \begin{tabular}[t]{>{\bf}l<{} @{\ } | l c c c H c c H c H c c c}
        &
        \rotl{Methodology}
        & \rotlsp{
          State\;
          \parbox[b]{2.2cm}{
          \normalfont
          \stglobal{}  global, \stlocal{}  local \\
        \stfo{}  first-order, \stho{}  HO
        }
        }
        & \rotc{Higher-order functions}
        &
          \rotc{\DP definition}
        &
        & \rotc{Verified}
        & \rotc{Beyond comp.}
        &
        & \rotc{Adaptive}
        &
        & \rotc{Interactive}
        \\
        \hline
        \arrayrulecolor{gray}
        Fuzz \cite{fuzz} & lin. types & \no & \yes & \(\err\) & \old & \no & \no & seq & \no & \yes & \no \\
        \apfuzz \cite{deAmorim:Probabilistic:2021} & lin. typ., path adj. & \no & \yes & \appdp & \no & \no & \no & seq & \no & \no & \no \\
        DFuzz \cite{dfuzz} & lin. dep. types & \no & \yes & \(\err\) & \old & \no & \no & seq & \no & \yes & \no \\
        AFuzz \cite{afuzz} & dyn. typ. + priv. filter & \no & \yes & \appdp & \old & \no & \no & adapt. & \yes & \no & \no \\
        Fuzzi \cite{fuzzi} & Fuzz-like + apRHL & \no & \no & \appdp & \yes & \no & \yes & seq,adv & \no & \yes & \no  \\
        Duet \cite{duet} & sens. + priv. types & \no
        & \yes & \appdp, RDP, (z/t)CDP & \yes & \no & \no & seq & \no & \no & \no \\
        Jazz \cite{jazz} & ctxt'l sens. + priv. typ. & \no & \yes & \appdp, RDP, zCDP & \yes & \no & \no & seq & \no & \no & \no \\
        DPella \cite{dpella} & eDSL dep. types & \stmonad & \yes & \(\err\) & \yes & \no & \no & seq, par & \no & \yes & \no
        \\
        Solo \cite{solo} & eDSL dep. types & \stmonad & \yes & \appdp, RDP & \yes & \no & \no & seq, adv & \no & \no & \no \\
        Spar \cite{param-sens} & eDSL dep. types & \stmonad & \yes & \(\err\) & ? & \no & \no
        & seq
        & \no & \no & \no \\
        \HOARe \cite{Barthe:HigherOrder:2015} & rel. refinement types & \no & \yes & \appdp & \yes & \no & \no & seq & \no & \yes & \no \\
        LightDP \cite{lightdp} & types + annot. & \stglobal\; \stfo & \no & \(\err\) & \yes & \no & \yes & seq & \no & \yes & \no \\
        \hline
        SampCert \cite{sampcert} & semantic & \no & \yes & \appdp, {\normalfont zCDP} & \yes & \yes & \yes
        & seq
        & \no & \yes & \no \\
        DP/Isabelle \cite{Sato:Formalization:2025} & semantic & \no & \no & \appdp & \yes & \yes & \yes
        & seq
        & \no & \no & \no \\
        \hline
        apRHL+ \cite{apRHL,relational_via_prob_coupling,advanced_for_diff_privacy} & rel. prob. Hoare logic & \stglobal\; \stfo & \no & \appdp & \old & \no & \yes & adv & \no & \yes & \iadv \\
        HO-RPL \cite{HORHL} & rel. prob. Hoare logic & \stglobal\; \stfo & \yes & \appdp & \no & \no & \no
        & seq
        & \no & \no & \iadv \\
        \thelogic& rel. prob. sep. logic & \stlocal\; \stho  & \yes & \appdp & \yes & \yes & \yes
        & seq/\(\expect \aerr\)
        & \yes & \yes & \yes
      \end{tabular}
    }
    \caption{Comparison of \DP systems. ``Beyond comp.'' =  support for mechanisms whose \DP goes beyond composition laws (\eg, SVT, RNM).
    }
    \label{fig:related}
  \end{figure}

\paragraph{Types for \DP}
A wide range of type systems ensuring \DP have been developed.
Fuzz~\cite{fuzz} and its variants track function sensitivity via linear types and rely on metric composition to statically ensure pure \DP for a probabilistic \(\uplambda\)-calculus. DFuzz~\cite{dfuzz} integrated linear and dependent types to improve sensitivity analysis. \apfuzz~\cite{deAmorim:Probabilistic:2021} extends Fuzz with support for \((\err,\aerr)\)-\DP.
The two-level type system of Adaptive Fuzz~\cite{afuzz} enhances static typing by integrating a trusted privacy filter into the language runtime. The system allows programming with adaptive composition by dynamically type-checking programs during execution and composing them according to the privacy filter. In \thelogic, privacy filters are just regular programs that can be verified in their own right.

Duet~\cite{duet} is a linear type system supporting various notions of \DP by a separation of the language into a sensitivity and a privacy layer which interact through bespoke composition laws that restrict rescaling, which limits the kinds of higher-order functions that can be type-checked.
Jazz~\cite{jazz} lifts some of these restrictions by introducing latent contextual effect types.
\HOARe{}~\cite{Barthe:HigherOrder:2015} encodes sensitivity and privacy information in relational refinement types for a pure calculus.

The DPella system~\cite{dpella}, Solo~\cite{solo}, and Spar~\cite{param-sens} leverage extensions to Haskell's type system to encode sensitivity (or distance) information via dependent types instead of linear types.

None of these systems support %
mutable state, or the verification of programs whose privacy requires advanced budget management instead of following directly from composition laws.

LightDP~\cite{lightdp} employs a dependent relational type system to bound distances in program variables. With SMT-backed type inference, LightDP can verify \DP for some mechanisms beyond composition laws (\eg, non-interactive SVT), but the method does not extend to advanced language features or flexible privacy budget analysis as used, \eg, in Report Noisy Max.

\paragraph{Logics for \DP}
The apRHL(+) program logics~\cite{apRHL,relational_via_prob_coupling,advanced_for_diff_privacy}  can prove \((\err,\aerr)\)-\DP for programs written in a first-order while-language.
Reasoning about basic mechanisms is well supported in apRHL, and it has been applied to advanced mechanisms such as SVT. Verification of interactive \DP is supported through a specialized rule for adversaries that can interact with mechanisms through fixed patterns.
In apRHL, adversaries are programs subject to a number of side-conditions; in contrast, our model of interactive computations via higher-order functions does not come with syntactic restrictions and can be composed modularly.
For instance, we can recover the apRHL model of SVT used by \citet[Fig.~1]{advanced_for_diff_privacy} through our streaming SVT (\cref{sec:svt-stream}) by instantiating the stream of queries \(\vqstream\) with the adversary \(\mathcal{A}\) of \loccit.
The privacy budget in apRHL is handled via a grading on judgments which offers less flexibility compared to \thelogic's privacy credits.
On the other hand, working with a concrete denotational model allowed the authors of \cite{advanced_for_diff_privacy} to extend apRHL with support for ``advanced composition'' in the sense of \cite[\S3.5.2]{Dwork:Algorithmic:2014} and derive the corresponding bounds for SVT, which is not possible in \thelogic.

A variant of the EasyCrypt prover supports apRHL but the implementation of the apRHL rules are part of the trusted code base, whereas the rules of \thelogic are proven sound in a proof assistant.
During the development of our case studies we found a bug in the Laplace sampling rule of EasyCrypt\footnote{We have disclosed the bug to the EasyCrypt team.} which has accidentally been exploited in a user-contributed privacy proof of Report Noisy Max.
Our foundational approach would have prevented us from introducing such an erroneous rule.
Proof search for coupling-based proofs of \DP for first-order programs was studied in \cite{Albarghouthi:Synthesizing:2017}. Their definition of choice couplings inspired our \rref{Laplace-choice} rule.

Fuzzi~\cite{fuzzi} integrates an apRHL-style logic with a Fuzz-inspired sensitivity- and privacy-logic; working at the intersection of the languages of the two systems, it does not support mutable state or higher-order functions.

The HO-RPL logic extends the ideas from apRHL to support higher-order functions and continuous distributions  by giving a denotational semantics of programs in Quasi-Borel Spaces, but it is not known how to extend this approach to other language features that our challenge problems require such as dynamic allocation or higher-order store.

The Isabelle/HOL formalization of \DP~\cite{Sato:Formalization:2025} develops the mathematical theory of \DP in the continuous setting.
Working directly in a measure-theoretic semantics, they prove privacy of Report Noisy Max, stating however that ``the formal proof is quite long''. It is unclear how to scale their approach to a language like \thelang.

The SampCert project \cite{sampcert} formalized \DP in the Lean prover by interpreting a shallow embedding of a while-like language in the same kind of denotational semantics that apRHL is based on.
Rather than assuming that the language has a primitive that samples from the Laplacian, SampCert proves that an efficient implementation of a sampler realizes the Laplace distribution.
Their approach is particularly well-suited to carrying out the low-level probabilistic reasoning and focuses less on building modular systems.
SampCert formalizes non-interactive variants of AT and SVT.

\paragraph{Separation logic}
Several probabilistic separation logics exist, but only \thelogic supports reasoning about \DP.
The Approxis~\cite{approxis} relational separation logic supports reasoning about approximate program equivalence. Their notion of \(\alpha\)-approximate equivalence amounts to \((0, \alpha)\)-\dipr in our setting, \ie, our additive privacy credits correspond to their ``error credits''. Approxis can prove cryptographic security or correctness of samplers but cannot express \DP.
Approxis includes specific provisions for reasoning about samplers such as the \textsc{spec-coupl-err-1} clause and ``expectation-preserving composition'' that are not present in our model. Approxis also develops a logical relation model for contextual equivalence.
We believe that \thelogic can be extended to support these features, and effectively subsume Approxis for additive error credits. Whether a expectation-preserving composition law can be formulated for the full generality of Clutch-DP remains a question for future work.
Bluebell~\cite{bluebell} encodes coupling-based relational reasoning via a conditioning modality, but only supports exact program equivalences for terminating first-order programs.

\paragraph{(Non-) Termination}
When verifying \DP, we have to decide whether non-termination is considered an observable behavior that can leak information about the presence of an individual in a database.
This is a well-known consideration in the context of \DP (\eg, \cite[\S3.5]{fuzz}, \cite[p.11]{apRHL}), and the most common solution to this problem is to rule out non-terminating programs a priori, \eg, by restricting the programming language syntactically, by including side-conditions in the program logic, or simply by not supporting reasoning about not-obviously-terminating programs.
\thelogic is a partial correctness logic: it neither assumes nor guarantees termination, and instead follows a more flexible approach by allowing verification of \DP for an arbitrary \thelang program \(\vf\). If we separately also prove that \(\vf\) is terminating on all inputs, we recover termination-sensitive \DP.
Standard techniques for ensuring termination of probabilistic programs apply \cite{termination-book,termination-rules}, and termination for \thelang-like languages
can be handled via separation logics for reasoning about termination probabilities \cite{eris, caliper} or expected runtime \cite{tachis, lohse2024iris}.

%% file: conclusion.tex
\section{Conclusion and Future Work}
\label{sec:conclusion-future-work}

We have developed \thelogic, a probabilistic higher-order separation logic for
\DP. To demonstrate how \thelogic enables modular verification
of \DP libraries, we addressed three representative challenges and verified a wide
range of case studies involving interactive mechanisms, privacy filters, and
memoization. \thelogic is proven sound as a library for Iris in the \rocq Prover.

In future work, we would like to extend \thelogic to model concurrency to reason about local \DP in a distributed setting.
\thelogic supports reasoning about generic mechanisms and about Laplace distributions, but it should be possible to extend the operational semantics with, say, Gaussian distributions. This should not invalidate any rules of the logic since Hoare quadruples are defined for a general probabilistic operational semantics and our key lemmas such as choice-composition are valid for arbitrary distributions.
To prove approximate \DP of programs based on the Gaussian, new rules would be required.
A challenging problem would be to extend \thelogic with support for further reasoning principles such as advanced composition in the sense of \DP or privacy amplification by subsampling \cite{dl-dp,subsampling-analyses}.
We would also like to integrate verified sampling mechanisms as developed by, \eg{}, \citet{sampcert}, to provide end-to-end \DP guarantees by adapting the techniques of \citet{eris} to verify rejection samplers to the relational setting.
Finally, it would be interesting to integrate other divergences~\cite{Sato:Approximate:2019} with relational separation logics to model, \eg, Rényi-\DP.

%% file: epilogue.tex
\section*{Data Availability Statement}
The Rocq formalization accompanying this work is available on Zenodo \cite{zenodo:clutch-dp} and
on GitHub at %
\href{https://github.com/logsem/clutch/tree/pldi26-clutch-dp}{github.com/logsem/clutch/tree/pldi26-clutch-dp}.

\begin{acks}
  This work was supported in part by the \grantsponsor{NSF}{National Science Foundation}{}, grant no.~\grantnum{NSF}{2338317}, the \grantsponsor{Carlsberg Foundation}{Carlsberg Foundation}{}, grant no.~\grantnum{Carlsberg Foundation}{CF23-0791}, \grantsponsor{Villum}{Villum}{} Investigator grants, no. \grantnum{Villum}{VIL25804} and no. \grantnum{Villum}{VIL73403}, Center for Basic Research in Program Verification (CPV), from the VILLUM Foundation, and the European Union (\grantsponsor{ERC}{ERC}{}, CHORDS, \grantnum{ERC}{101096090}).
  Views and opinions expressed are however those of the author(s) only and do not necessarily reflect those of the European Union or the European Research Council.
  Neither the European Union nor the granting authority can be held responsible for them.
\end{acks}

%% file: appendix.tex
\section{Semantic Model}
\label{sec:app-model}

In this section we dive into the model behind \thelogic{} and show how our adequacy theorem is proven.
We begin by recalling some notions of probability theory.

\subsection{Probability Theory and Probabilistic Couplings}

Given a countable set $A$, a probability (sub-)distribution $\mu$ over $A$ is a
function $\mu \colon A \to [0,1]$ such that $\sum_a \mu(a) \leq 1$. We use the
notation $\Distr{A}$ to denote the set of distributions over $A$. Distributions
have a well-known monadic structure, see, \eg, \cite{approxis}. We use $\mret{a}$
and $\mu\mbindi f$ to denote the return and bind operations. Given a random
variable $f \colon A \to [0,1]$, its expected value is given by $\expect[\mu]{f} \triangleq \sum_a \mu(a) f(a)$.

We recall the notion of $(\varepsilon,\delta)$-approximate coupling, due to \cite{advanced_for_diff_privacy, Sato:Approximate:2016}:

\begin{definition}

Let $A, B$ be countable types, and $\Phi \subseteq A \times B$ a relation. Given
two real-valued parameters $0 \leq \varepsilon, \delta$, we say that there is
an $(\varepsilon, \delta)$-approximate $\Phi-$coupling between distributions
$\mu_1 \colon \Distr{A}, \mu_2 \colon \Distr{B}$ if, for any real-valued random
variables $f : A \to [0,1], g : B \to [0,1]$ such that $\forall (a,b)\in\Phi, f a \leq g b$,
the following holds : $\expect[\mu_1]{f} \leq \exp(\varepsilon) \cdot \expect[\mu_2]{g} + \delta$.
We denote the existence of such a coupling by $\dpcoupl{\mu_1}{\mu_2}{\varepsilon}{\delta}{\Phi}$.

\end{definition}

The reason we are interested in approximate couplings is that there is a tight connection between
them and \DP. Indeed, when instantiating $\Phi$ to be the equality relation
we have the following result:

\begin{theorem}
Let \(f : \DB \to \Distr X\) be a function over databases. Then, $f$ is \((\varepsilon,\delta)\)-\dipr
iff for any inputs $b,b'\colon \DB$ such that \(\dr[\DB] b b' 1\), we have $\dpcoupl{f~b}{f~b'}{\varepsilon}{\delta}{(=)}$.
\end{theorem}

Approximate couplings are the key component of our relational program logic.
One can think of $\mu_1,\mu_2$ as two randomized computations, $\Phi$ as a
relational postcondition we wish to establish and $(\varepsilon,\delta)$ as a
privacy budget that we can spend in order to prove the postcondition. We will see
later in the section how to interpret $\varepsilon$ and $\delta$ as separation logic
resources, but for now notice that they have an affine flavor to them, since they satisfy
the following monotonicity lemma:

\begin{lemma}
  Let $\mu_1 \colon \Distr{A}, \mu_2 \colon \Distr{B}$, $\Phi\subseteq A\times B$.
  Assume that $\dpcoupl{\mu_1}{\mu_2}{\varepsilon}{\delta}{\Phi}$. Then, for all
  $\varepsilon',\delta',\Phi'$ such that $\varepsilon \leq \varepsilon'$, $\delta \leq \delta'$,
  $\Phi \subseteq \Phi'$, we also have $\dpcoupl{\mu_1}{\mu_2}{\varepsilon'}{\delta'}{\Phi'}$.
\end{lemma}

In order to build a logic around couplings, we need them to be able to compose them along the operations
of the underlying distribution monad. The following results are well-known:

\begin{theorem}
The distribution monad operations lift to couplings, in the sense that:
\begin{itemize}
	\item Let $a \colon A, b \colon B$ and $\Phi \subseteq A \times B$ such that
        $(a,b) \in \Phi$. Then $\dpcoupl{\mret{a}}{\mret{b}}{\varepsilon}{\delta}{\Phi}$.

        \item Let $\mu_1 \colon \Distr{A}, \mu_2 \colon \Distr{B}, f \colon A \to \Distr{A'}, g \colon B \to \Distr{B'}$
	and $\Phi \subseteq A\times B, \Psi \subseteq A'\times B'$
	such that $\dpcoupl{\mu_1}{\mu_2}{\varepsilon}{\delta}{\Phi}$ and,
	for all $(a,b)\in\Phi$, $\dpcoupl{f a}{g b}{\varepsilon'}{\delta'}{\Psi}$.
		Then, $\dpcoupl{(\mu_1 \mbindi f)}{(\mu_2 \mbindi g)}{\varepsilon+\varepsilon'}{\delta+\delta'}{\Psi}$.
\end{itemize}
\end{theorem}

In order to support a more flexible form of composition, such as in our Above Threshold example (\cref{sec:at}), we introduce
a more general version of the composition lemma above, where the choice of coupling in the continuation
can depend on the result of the first step. This is heavily inspired by choice couplings \cite{advanced_for_diff_privacy},
but it really shines in our setting, since we have a resourceful treatment of the privacy budget.

\begin{theorem}\label{thm:choice-bind}
  Let $\mu_1 \colon \Distr{A}, \mu_2 \colon \Distr{B}, f \colon A \to \Distr{A'}, g \colon B \to \Distr{B'}$.
  Assume we have a predicate $\Xi \subseteq A$, and $\Phi_1, \Phi_2 \subseteq A\times B, \Psi \subseteq A'\times B'$,
  with $\Phi_1, \Phi_2$ disjoint in the sense that, $\forall a, a', b. a \in \Xi \wedge a' \not\in \Xi \rightarrow ((a,b)\not\in \Phi_1 \vee (a',b)\not\in \Phi_2)$.
  Assume further that:
  \begin{itemize}
  \item $\dpcoupl{\mu_1}{\mu_2}{\varepsilon_1}{\delta_1}{\Phi_1}$
  \item $\dpcoupl{\mu_1}{\mu_2}{\varepsilon_2}{\delta_2}{\Phi_2}$
  \item For all $a,b$ such that $a \in \Xi$ and $(a,b)\in\Phi_1$, $\dpcoupl{f a}{g b}{\varepsilon_1'}{\delta_1'}{\Psi}$.
  \item For all $a,b$ such that $a \not\in \Xi$ and $(a,b)\in\Phi_2$, $\dpcoupl{f a}{g b}{\varepsilon_2'}{\delta_2'}{\Psi}$.
  \end{itemize}

	Then, $\dpcoupl{(\mu_1 \mbindi f)}{(\mu_2 \mbindi g)}{\varepsilon}{\delta}{\Psi}$, where
  $\varepsilon \triangleq {\sf max}(\varepsilon_1 + \varepsilon_1', \varepsilon_2 + \varepsilon_2')$
  and
  $\delta \triangleq \delta_1 + \delta_2 + {\sf max}(\delta_1', \delta_2')$

\end{theorem}

The idea behind the statement above is that it allows us to use two different
couplings for the first step: one over $\Phi_1$ and another over $\Phi_2$. Then
get a sample $a$ from $\mu_1$, and we choose which of the couplings to use
depending on whether $a$ lands in $\Xi$, and we continue the rest of the
execution. In particular, this allows us to optimize the use of the $\varepsilon$
component of the privacy budget, since we can set the amount to use in each of the
variants and only spend what we need. In the case of the $\delta$ component, we
can also set the optimal amount for the continuation, but not for the first step,
both $\delta_1$ and $\delta_2$ need to be spent no matter the result.

\subsection{Hoare Quadruples and the Weakest Precondition}

Hoare Quadruples are not a primitive notion in \thelogic.
Instead, they are defined in terms of a weakest precondition predicate (WP).
This is a unary predicate about the left-hand side program.
The right-hand program \(\expr_R\) is tracked by a resource \(\spec(\expr_R)\).
\[
  \hoareref {\prop} {\expr} {\expr'} {v\ v'\,.\,\propB}
  \quad\eqdef\quad
  \always
  (\prop \wand
  \Exists \lctx . \spec(\fillctx \lctx[\expr']) \wand \wpre {\expr} {\val . \Exists \val' . \spec(\fillctx \lctx[\val']) \sep \propB(\val,\val')})
\]
to state that \(\expr\) refines \(\expr'\) and that if \(\expr\) terminates with \(\val\) then  \(\expr'\) terminates with \(\val'\) and the postcondition \(\prop\) is satisfied.

The WP, whose definition is shown below, couples the execution of the
implementation program together with the execution of the specification program
and ensures that the postcondition holds at the end. The construction is
similar to the one presented in Approxis~\cite{approxis}, but adapted to our
more general notion of coupling. Crucially, we also employ an updated program
coupling modality to support choice coupling composition at the level of
program steps. Notice that this definition tracks the two components of the
error, both $\varepsilon$ and $\delta$. As usual, the weakest precondition is a
predicate defined as a guarded fixpoint, whose existence is ensured by the
presence of the later modality ($\later$) in front of the recursive occurrence
of the weakest precondition:
\begin{align*}
  \wpre{\expr_{1}}[\ghostcode{\mask}]{\pred} \eqdef{}
  & \All \state_{1}, \cfg_{1}', \err_{1}, \delta_1 .
    \stateinterp(\state_{1}, \cfg_{1}', \err_{1}, \delta_1) \wand \\
  & \quad \ghostcode{\pvs[\mask][\emptyset]} \specCoupll{\state_{1}}{\cfg_{1}'}{\err_{1}}{\delta_1}\spac\{ \state_{2}, \cfg_{2}', \err_{2},\delta_2 \ldotp \\
  & \qquad \big(\expr_{1} \in \Val \sep
    \ghostcode{\pvs[\emptyset][\mask]} \stateinterp(\state_{2}, \cfg_{2}', \err_{2}, \delta_2) \sep \pred(\expr_{1})\big) \lor{} \\
  & \qquad
    \big( \expr_{1} \not\in \Val \sep
    \progCoupll{(\expr_{1}, \state_{2})}{\cfg_{2}'}{\err_{2}}{\delta_2}
    \spac\{ \expr_{2}, \state_{3}, \cfg_{3}', \err_{3}, \delta_3 \ldotp  \\
  & \quad\qquad \later \specCoupll{\state_{3}}{\cfg_{3}'}{\err_{3}}{\delta_3}
    \spac\{ \state_{4}, \cfg_{4}', \err_{4}, \delta_4 \ldotp
    \ghostcode{\pvs[\emptyset][\mask]} \stateinterp(\state_{4}, \cfg_{4}', \err_{4}, \delta_4) \sep \wpre{\expr_{2}}[\ghostcode{\mask}]{\pred} \} \} \big) \}
\end{align*}

The weakest precondition requires us to own the \emph{state interpretation}
$\stateinterp(\state_{1}, \cfg_{1}', \err_{1}, \delta_1)$, which is a predicate
that maps the different resources in \thelogic{} (the physical state of the
implementation program, the configuration of the specification program, and the
privacy budget) to their corresponding resource algebras. We then have to prove
two different cases, depending on whether the current program $e$ is or is not
a value. In the former case, we can just give back the state interpretation and
prove $\pred$. Otherwise, we have to prove a property about the distribution
that results after taking one program step. This is phrased in terms of the
composition of two modalities, the \emph{specification coupling modality} and
the \emph{program coupling modality}.

The specification coupling modality, initially introduced in Approxis, accounts
for logical updates to the current state of the execution. In particular, it
allows for the specification program to take physical steps. It is defined as
and inductive predicate given by the following rules:
\begin{mathpar}
  \inferH{spec-coupl-ret}
  { \Phi(\state, \cfg', \err, \delta) }
  { \specCoupl{\state}{\cfg'}{\err}{\delta}{\Phi} }
  \and
  \inferH{spec-coupl-bind}
  {
    \dpcoupl{\mu_1}{(\mu_1' \mbindi \Lam \state_2' . \pexec_n(\expr_1', \state_2'))}{\err_1}{\delta_1}{R} \\
    \err_1 + \err_2 \leq \err \\
    \delta_1 + \delta_2 \leq \delta \\
    \erasable(\mu_1, \state_1) \\
    \erasable(\mu_1', \state_1') \\
	\All (\sigma_2, \cfg_2') \in R . \ghostcode{\pvs[\emptyset]} \specCoupl{\state_2}{\cfg_2'}{\err_2}{\delta_2}{\Phi}
  }
  { \specCoupl{\state_1}{(\expr_1', \state_1')}{\err}{\delta}{\Phi} }
\end{mathpar}

For the purpose of this work, the notion of erasability can mostly be ignored, this is
used to support presampling tapes~\cite{clutch}. By the \rref{spec-coupl-ret}
rule, if the current implementation and specification programs, and error
budget satisfy $\Phi$, we can immediately conclude
$\specCoupl{\state}{\cfg'}{\err}{\delta}{\Phi}$. The \rref{spec-coupl-bind}
considers the recursive case. We can split our current credit budget $\err,\delta$
into $\err_1+\err_2, \delta_1+\delta_2$ and use $\err_1,\delta_1$ to take a
number $n$ of steps on the specification program under an approximate coupling
that satisfies an intermediate relation $R$. Then from $(\sigma_2, \cfg_2') \in R$,
we continue to recursively stablish $\specCoupl{\state_2}{\cfg_2'}{\err_2}{\delta_2}{\Phi}$,
with our remaining credit budget $\err_2,\delta_2$.

The program coupling modality couples exactly one physical step of the
left-hand side program with an arbitrary number of physical steps $n$ of
right-hand side program. In addition, the right-hand side program is also
allowed to execute an erasable update before its physical steps. Choice
couplings are baked into the definition of the program coupling modality, thus
the intermediate pair of states that the program reach can satisfy one of
either $R_1$ or $R_2$ before executing the continuation. The amount of error
credits that is available for the continuation may also depend on this choice.

\begin{mathpar}
  \inferH{prog-coupl-bind}
	{
		\dpcoupl{\stepdistr(\expr_1, \state_1)}{(\mu_1' \mbindi \Lam \state_2' . \pexec_{n}(\expr_1', \state_2'))}{\err_1}{\delta_1}{R_1} \\
		\dpcoupl{\stepdistr(\expr_1, \state_1)}{(\mu_1' \mbindi \Lam \state_2' . \pexec_{n}(\expr_1', \state_2'))}{\err_2}{\delta_2}{R_2} \\
    \red(\expr_1, \state_1) \\
    \err_1 + \err_2 \leq \err \\
    \err_1' + \err_2' \leq \err \\
    \delta_1 + \delta_2 + {\sf max}(\delta_1' + \delta_2') \leq \delta \\
	\erasable(\mu_1', \state_1') \\\\ \\\\
    \All ((\expr_2, \sigma_2), (\expr_2', \sigma_2')).
	((P (\expr_2, \sigma_2) \wedge R_1((\expr_2, \sigma_2), (\expr_2', \sigma_2'))) \ghostcode{\pvs[\emptyset]} \Phi(\expr_2, \state_2, \expr_2', \state_2', \err_1',\delta_1')) \ast \\\\
	((\neg P (\expr_2, \sigma_2) \wedge R_2((\expr_2, \sigma_2), (\expr_2', \sigma_2'))) \ghostcode{\pvs[\emptyset]} \Phi(\expr_2, \state_2, \expr_2', \state_2', \err_2',\delta_2'))
    }
	{\progCoupl{(\expr_1, \state_1)}{(\expr_1', \state_1')}{\err}{\delta}{\Phi}}
\end{mathpar}

\subsection{An Adequacy Theorem}

With these definitions in hand, we can state the adequacy theorem of \thelogic,
now in full generality:
\begin{theorem}
\label{thm:adeq-gen}
  Let $e,e'$ be two \thelang functions and $\Psi \colon \Val \times \Val \to \Prop$
  a proposition over (final) values.
	If $\spec(e') \ast \uptom{\varepsilon} \ast \upto{\delta} \wand \wpre{e}{v, v'. \spec(v') \ast \Psi}$
	holds in \thelogic{}, then, for any initial states $\sigma,\sigma'$, we have $\dpcoupl{\exec(e,\sigma)}{\exec(e',\sigma')}{\varepsilon}{\delta}{\Psi}$.
\end{theorem}

The proof is done by induction on the execution of the left-hand side program, by proving the
lemma below, and finally taking a limit:
\begin{lemma}
  Let $e,e'$ be two \thelang functions and $\Psi \colon \Val \times \Val \to \Prop$
  a proposition over (final) values.
	If $\spec(e') \ast \uptom{\varepsilon} \ast \upto{\delta} \wand \wpre{e}{v, v'. \spec(e') \ast \Psi}$
	holds in \thelogic{}, then, for any $n\in\nat$ and any initial states $\sigma,\sigma'$, we have
	$\dpcoupl{\exec_{n}(e,\sigma)}{\exec(e',\sigma')}{\varepsilon}{\delta}{\Psi}$.
\end{lemma}

The key part of the proof happens when the program on the left is not a value,
and has to take a step. This requires two applications of
\rref{spec-coupl-bind} and one application of \rref{prog-coupl-bind} to
establish the specification coupling in the non-value case of the WP. In
particular, \rref{prog-coupl-bind} allows us to advance the program on the left
exactly one step, which lets us use our inductive hypothesis about
$\exec_{n}(e,-)$ to reason about $\exec_{n+1}(e,\sigma)$.

\section{Case Study: Report Noisy Max}
\label{sec:report-noisy-max}

The Report Noisy Max mechanism (RNM) is a \DP mechanism that receives a list of
$N$ 1-sensitive queries, runs them privately by adding noise sampled from
$\Laplace\ (\verr/2)$, and returns the index of the query with the highest
value. While a na\"ive analysis of the algorithm using composition theorems
would show that the algorithm is $(N\cdot\verr/2)$-\dipr, the only information
that is released is the index of the highest query and all other (noisy) results are discarded. This suggests that a
better analysis should be possible. In fact, one can prove that RNM is
$\verr$-\dipr, in other words, its privacy cost is constant and is completely independent of the number
of queries.

\newcommand\maxI{\langv{maxI}}
\newcommand\maxA{\langv{maxA}}
\newcommand\vxs{\langv{xs}}
\newcommand\vxstapes{\langv{xs\_tapes}}
\newcommand\vnoisyxs{\langv{noisy\_xs}}
We now show how this is proven in \thelogic. In this example, we make use of \emph{presampling tapes}, a reasoning principle for talking about future probabilistic outcomes~\cite{clutch,approxis}. Our implementation of RNM (see \cref{fig:rnm-impl}) therefore utilizes code to manipulate tapes, and one can show that this implementation is contextually equivalent to one with all the tape annotations removed~\cite{clutch}.

\begin{figure}[htbp]
  \begin{tabular}[t]{>{$}l<{$} @{} >{$}l<{$} @{} >{$}l<{$} }
    \DumbLet\,\, &\RNM\ \vq\ \vN\ \vdb = \\
                 &\Let \vxs = \listinit\ \vN\ (\Lam \vi.\vq\ \vi\ \vdb) in\\
                 & \Let \vxstapes = \listmap\ (\Lam \vx. (\vx, \AllocTapeLaplace\ (\verr/2)\ \vx))\ \vxs in\\
                 & \Let \vnoisyxs = \listmap\ (\Lam (\vx, \lbl). \Laplace\ (\verr/2)\ \vx\ \lbl)\ \vxstapes in\\
                 & \listmaxindex\ \vnoisyxs
    \end{tabular}
\caption{The Report Noisy Max mechanism}
\label{fig:rnm-impl}
\end{figure}

The implementation of RNM takes three arguments: the query function $\vq$, the number of queries $\vN$, and the database $\vdb$.
The first line of RNM computes the (exact) results of running the query function for all indices $0\leq\vi<\vN$ on the database and stores the result in the list \(\vxs\).
The second line pairs each result \(\vx\) with a freshly allocated presampling tape label for a Laplacian with spread \(\verr/2\) and mean \(\vx\).
This ghost code will later allow us to reason about the Laplacian samples.
On line three, we compute the noisy results \(\vnoisyxs\) by adding Laplace noise, and finally on line four we return the index of the largest noisy value.

We prove that RNM is internally differentially private for any 1-sensitive family of queries \(\vq\). Concretely, this amounts to
\begin{align}
\label{eq:rnm-spec}
  \All\ \vq\ \vN\ \vdb\ \vdb'.
  \; (\All i. \sens{1}{\vq\ i}) \;
  \wand
  \;
  \hoareVref{\adj{\vdb}{\vdb'} \sep \uptom{\verr}}
  {\RNM\ \vq\ \vN\ \vdb}
  {\RNM\ \vq\ \vN\ \vdb'}
  {\val\ \val'\,.\, \val = \val' }
\end{align}
By soundness of \thelogic (\cref{thm:soundness-dp}), \(\RNM\ \vq\ \vN\) is \(\verr\)-\dipr for suitable queries~\(\vq\) and for all \(\vN\).

The key idea in this proof is to utilize tapes to perform a relatively complicated presampling operation on all of the \(\vN\) allocated Laplace tapes at the logic level \emph{in one logical step}.
This ensures that the index of the highest presampled result is the same for \(\vdb\) and for \(\vdb'\). Then, when the \(\Laplace\) instructions on line three are executed, we \emph{deterministically} read out the presampled values from the tapes, and establish the postcondition directly.

Let us step through the proof line-by-line. After querying the database at the beginning, we obtain two lists of results represented by lists $l_1$ and $l_1'$ for the program and specification side, respectively.
Since the queries are 1-sensitive, it follows that $l_1$ and $l_1'$ are pairwise-adjacent.
Next, we allocate tapes for these results,  obtaining a list of tape labels ${\iota}s$ and ${\iota}s'$ where all of them point to empty tapes:
\begin{mathpar}
    \Sep_{(x, \lbl)\in l_1;{\iota}s}  \proglaplacetape{\lbl}{\verr/2}{x}{[]} %
  \quad\sep
    \Sep_{(x', \lbl')\in l_1';{\iota}s'} \speclaplacetape{\lbl'}{\verr/2}{x'}{[]}%

  \end{mathpar}

  We then perform a logic-level resource update, where we perform a presampling operation on the tapes in one go (similar to {\TirNameStyle{\normalsize rel-couple-tapes}} from the Clutch paper~\cite{clutch}, albeit being more complicated), which consumes $\uptom{\verr}$ from the precondition.
  After this operation, the program and specification tapes contain $l_2$ and $l_2'$, respectively, where the index of the maximum element in $l_2$ and $l_2'$ is the same. Instead of consuming $\uptom{N\cdot\verr}$ to apply \rref{Laplace-shift} $N$ times for each of the $N$ samplings, roughly speaking, this operation enables us to consume exactly $\uptom{\verr}$ to pay for the component of the $\Laplace$ coupling which returns the largest result among the list of samplings.
  This operation is sound in \thelogic by leveraging a pointwise-equality argument, also used in \citet{Hsu:Probabilistic:2017}, and we obtain the following:
  \begin{equation*}
    \Sep_{(x, y,\lbl)\in l_1;l_2;{\iota}s}  \proglaplacetape{\lbl}{\verr/2}{x}{[y]} %
  \quad\sep
    \Sep_{(x', y', \lbl')\in l_1';l_2';{\iota}s'} \speclaplacetape{\lbl'}{\verr/2}{x'}{[y']}%
    \;\;\sep\; \langv{max\_index}\ l_2=\langv{max\_index}\ l_2'
\end{equation*}

Calling $\Laplace$ on these pre-populated tapes then deterministically returns the value stored in the tape. As a result, $\vnoisyxs$ and $\vnoisyxs'$ contains $l_2$ and $l_2'$, respectively, and the index of the maximum element of both lists is the same, establishing the postcondition of the specification.

%% file: pldi26-differential-privacy.bbl

\begin{thebibliography}{52}


\ifx \showCODEN    \undefined \def \showCODEN     #1{\unskip}     \fi
\ifx \showDOI      \undefined \def \showDOI       #1{#1}\fi
\ifx \showISBNx    \undefined \def \showISBNx     #1{\unskip}     \fi
\ifx \showISBNxiii \undefined \def \showISBNxiii  #1{\unskip}     \fi
\ifx \showISSN     \undefined \def \showISSN      #1{\unskip}     \fi
\ifx \showLCCN     \undefined \def \showLCCN      #1{\unskip}     \fi
\ifx \shownote     \undefined \def \shownote      #1{#1}          \fi
\ifx \showarticletitle \undefined \def \showarticletitle #1{#1}   \fi
\ifx \showURL      \undefined \def \showURL       {\relax}        \fi
\providecommand\bibfield[2]{#2}
\providecommand\bibinfo[2]{#2}
\providecommand\natexlab[1]{#1}
\providecommand\showeprint[2][]{arXiv:#2}

\bibitem[Abadi et~al\mbox{.}(2016)]%
        {dl-dp}
\bibfield{author}{\bibinfo{person}{Martin Abadi}, \bibinfo{person}{Andy Chu},
  \bibinfo{person}{Ian Goodfellow}, \bibinfo{person}{H.~Brendan McMahan},
  \bibinfo{person}{Ilya Mironov}, \bibinfo{person}{Kunal Talwar}, {and}
  \bibinfo{person}{Li Zhang}.} \bibinfo{year}{2016}\natexlab{}.
\newblock \showarticletitle{Deep Learning with Differential Privacy}. In
  \bibinfo{booktitle}{\emph{Proceedings of the 2016 ACM SIGSAC Conference on
  Computer and Communications Security}} (Vienna, Austria)
  \emph{(\bibinfo{series}{CCS '16})}. \bibinfo{publisher}{Association for
  Computing Machinery}, \bibinfo{address}{New York, NY, USA},
  \bibinfo{pages}{308–318}.
\newblock
\showISBNx{9781450341394}
\urldef\tempurl%
\url{https://doi.org/10.1145/2976749.2978318}
\showDOI{\tempurl}


\bibitem[Abuah et~al\mbox{.}(2022)]%
        {solo}
\bibfield{author}{\bibinfo{person}{Chik{\'e} Abuah}, \bibinfo{person}{David
  Darais}, {and} \bibinfo{person}{Joseph~P. Near}.}
  \bibinfo{year}{2022}\natexlab{}.
\newblock \showarticletitle{Solo: A Lightweight Static Analysis for
  Differential Privacy}.
\newblock \bibinfo{journal}{\emph{Proc. ACM Program. Lang.}}
  \bibinfo{volume}{6}, \bibinfo{number}{OOPSLA2}, Article
  \bibinfo{articleno}{150} (\bibinfo{date}{Oct.} \bibinfo{year}{2022}).
\newblock
\urldef\tempurl%
\url{https://doi.org/10.1145/3563313}
\showDOI{\tempurl}


\bibitem[Aguirre et~al\mbox{.}(2021)]%
        {HORHL}
\bibfield{author}{\bibinfo{person}{Alejandro Aguirre}, \bibinfo{person}{Gilles
  Barthe}, \bibinfo{person}{Marco Gaboardi}, \bibinfo{person}{Deepak Garg},
  \bibinfo{person}{Shin-ya Katsumata}, {and} \bibinfo{person}{Tetsuya Sato}.}
  \bibinfo{year}{2021}\natexlab{}.
\newblock \showarticletitle{Higher-order probabilistic adversarial
  computations: categorical semantics and program logics}.
\newblock \bibinfo{journal}{\emph{Proc. ACM Program. Lang.}}
  \bibinfo{volume}{5}, \bibinfo{number}{ICFP}, Article \bibinfo{articleno}{93}
  (\bibinfo{date}{Aug} \bibinfo{year}{2021}), \bibinfo{numpages}{30}~pages.
\newblock
\urldef\tempurl%
\url{https://doi.org/10.1145/3473598}
\showDOI{\tempurl}


\bibitem[Aguirre et~al\mbox{.}(2024)]%
        {eris}
\bibfield{author}{\bibinfo{person}{Alejandro Aguirre},
  \bibinfo{person}{Philipp~G. Haselwarter}, \bibinfo{person}{Markus de
  Medeiros}, \bibinfo{person}{Kwing~Hei Li}, \bibinfo{person}{Simon~Oddershede
  Gregersen}, \bibinfo{person}{Joseph Tassarotti}, {and} \bibinfo{person}{Lars
  Birkedal}.} \bibinfo{year}{2024}\natexlab{}.
\newblock \showarticletitle{Error Credits: Resourceful Reasoning about Error
  Bounds for Higher-Order Probabilistic Programs}.
\newblock \bibinfo{journal}{\emph{Proc. ACM Program. Lang.}}
  \bibinfo{volume}{8}, \bibinfo{number}{ICFP}, Article \bibinfo{articleno}{246}
  (\bibinfo{date}{Aug} \bibinfo{year}{2024}), \bibinfo{numpages}{33}~pages.
\newblock
\urldef\tempurl%
\url{https://doi.org/10.1145/3674635}
\showDOI{\tempurl}


\bibitem[Albarghouthi and Hsu(2017)]%
        {Albarghouthi:Synthesizing:2017}
\bibfield{author}{\bibinfo{person}{Aws Albarghouthi} {and}
  \bibinfo{person}{Justin Hsu}.} \bibinfo{year}{2017}\natexlab{}.
\newblock \showarticletitle{Synthesizing Coupling Proofs of Differential
  Privacy}.
\newblock \bibinfo{journal}{\emph{Proceedings of the ACM on Programming
  Languages}} \bibinfo{volume}{2}, \bibinfo{number}{POPL} (\bibinfo{date}{Dec.}
  \bibinfo{year}{2017}), \bibinfo{pages}{58:1--58:30}.
\newblock
\urldef\tempurl%
\url{https://doi.org/10.1145/3158146}
\showDOI{\tempurl}


\bibitem[Balle et~al\mbox{.}(2018)]%
        {subsampling-analyses}
\bibfield{author}{\bibinfo{person}{Borja Balle}, \bibinfo{person}{Gilles
  Barthe}, {and} \bibinfo{person}{Marco Gaboardi}.}
  \bibinfo{year}{2018}\natexlab{}.
\newblock \showarticletitle{Privacy amplification by subsampling: tight
  analyses via couplings and divergences}. In
  \bibinfo{booktitle}{\emph{Proceedings of the 32nd International Conference on
  Neural Information Processing Systems}} (Montr\'{e}al, Canada)
  \emph{(\bibinfo{series}{NIPS'18})}. \bibinfo{publisher}{Curran Associates
  Inc.}, \bibinfo{address}{Red Hook, NY, USA}, \bibinfo{pages}{6280–6290}.
\newblock


\bibitem[Bao et~al\mbox{.}(2025)]%
        {bluebell}
\bibfield{author}{\bibinfo{person}{Jialu Bao}, \bibinfo{person}{Emanuele
  D'Osualdo}, {and} \bibinfo{person}{Azadeh Farzan}.}
  \bibinfo{year}{2025}\natexlab{}.
\newblock \showarticletitle{Bluebell: An Alliance of Relational Lifting and
  Independence for Probabilistic Reasoning}.
\newblock \bibinfo{journal}{\emph{Proc. ACM Program. Lang.}}
  \bibinfo{volume}{9}, \bibinfo{number}{POPL}, Article \bibinfo{articleno}{58}
  (\bibinfo{date}{Jan.} \bibinfo{year}{2025}), \bibinfo{numpages}{31}~pages.
\newblock
\urldef\tempurl%
\url{https://doi.org/10.1145/3704894}
\showDOI{\tempurl}


\bibitem[Barthe et~al\mbox{.}(2015a)]%
        {relational_via_prob_coupling}
\bibfield{author}{\bibinfo{person}{Gilles Barthe}, \bibinfo{person}{Thomas
  Espitau}, \bibinfo{person}{Benjamin Gr{\'{e}}goire}, \bibinfo{person}{Justin
  Hsu}, \bibinfo{person}{L{\'{e}}o Stefanesco}, {and}
  \bibinfo{person}{Pierre{-}Yves Strub}.} \bibinfo{year}{2015}\natexlab{a}.
\newblock \showarticletitle{Relational Reasoning via Probabilistic Coupling}.
  In \bibinfo{booktitle}{\emph{Logic for Programming, Artificial Intelligence,
  and Reasoning - 20th International Conference, {LPAR-20} 2015, Suva, Fiji,
  November 24-28, 2015, Proceedings}}.
\newblock
\urldef\tempurl%
\url{https://doi.org/10.1007/978-3-662-48899-7\_27}
\showDOI{\tempurl}


\bibitem[Barthe et~al\mbox{.}(2016a)]%
        {advanced_for_diff_privacy}
\bibfield{author}{\bibinfo{person}{Gilles Barthe},
  \bibinfo{person}{No{\'{e}}mie Fong}, \bibinfo{person}{Marco Gaboardi},
  \bibinfo{person}{Benjamin Gr{\'{e}}goire}, \bibinfo{person}{Justin Hsu},
  {and} \bibinfo{person}{Pierre{-}Yves Strub}.}
  \bibinfo{year}{2016}\natexlab{a}.
\newblock \showarticletitle{Advanced Probabilistic Couplings for Differential
  Privacy}. In \bibinfo{booktitle}{\emph{Proceedings of the 2016 {ACM} {SIGSAC}
  Conference on Computer and Communications Security, Vienna, Austria, October
  24-28, 2016}} \emph{(\bibinfo{series}{{{CCS}} '16})}.
  \bibinfo{pages}{55--67}.
\newblock
\urldef\tempurl%
\url{https://doi.org/10.1145/2976749.2978391}
\showDOI{\tempurl}


\bibitem[Barthe et~al\mbox{.}(2015b)]%
        {Barthe:HigherOrder:2015}
\bibfield{author}{\bibinfo{person}{Gilles Barthe}, \bibinfo{person}{Marco
  Gaboardi}, \bibinfo{person}{Emilio~Jes{\'u}s Gallego~Arias},
  \bibinfo{person}{Justin Hsu}, \bibinfo{person}{Aaron Roth}, {and}
  \bibinfo{person}{Pierre-Yves Strub}.} \bibinfo{year}{2015}\natexlab{b}.
\newblock \showarticletitle{Higher-{{Order Approximate Relational Refinement
  Types}} for {{Mechanism Design}} and {{Differential Privacy}}}. In
  \bibinfo{booktitle}{\emph{Proceedings of the 42nd {{Annual ACM SIGPLAN-SIGACT
  Symposium}} on {{Principles}} of {{Programming Languages}}}}
  \emph{(\bibinfo{series}{{{POPL}} '15})}. \bibinfo{publisher}{Association for
  Computing Machinery}, \bibinfo{address}{New York, NY, USA},
  \bibinfo{pages}{55--68}.
\newblock
\showISBNx{978-1-4503-3300-9}
\urldef\tempurl%
\url{https://doi.org/10.1145/2676726.2677000}
\showDOI{\tempurl}


\bibitem[Barthe et~al\mbox{.}(2016b)]%
        {apRHL+}
\bibfield{author}{\bibinfo{person}{Gilles Barthe}, \bibinfo{person}{Marco
  Gaboardi}, \bibinfo{person}{Benjamin Gr\'{e}goire}, \bibinfo{person}{Justin
  Hsu}, {and} \bibinfo{person}{Pierre-Yves Strub}.}
  \bibinfo{year}{2016}\natexlab{b}.
\newblock \showarticletitle{Proving Differential Privacy via Probabilistic
  Couplings}. In \bibinfo{booktitle}{\emph{Proceedings of the 31st Annual
  ACM/IEEE Symposium on Logic in Computer Science}} (New York, NY, USA)
  \emph{(\bibinfo{series}{LICS '16})}. \bibinfo{publisher}{Association for
  Computing Machinery}, \bibinfo{address}{New York, NY, USA},
  \bibinfo{pages}{749–758}.
\newblock
\showISBNx{9781450343916}
\urldef\tempurl%
\url{https://doi.org/10.1145/2933575.2934554}
\showDOI{\tempurl}


\bibitem[Barthe et~al\mbox{.}(2012)]%
        {apRHL}
\bibfield{author}{\bibinfo{person}{Gilles Barthe}, \bibinfo{person}{Boris
  K\"{o}pf}, \bibinfo{person}{Federico Olmedo}, {and} \bibinfo{person}{Santiago
  Zanella~B\'{e}guelin}.} \bibinfo{year}{2012}\natexlab{}.
\newblock \showarticletitle{Probabilistic relational reasoning for differential
  privacy}.
\newblock \bibinfo{journal}{\emph{SIGPLAN Not.}} \bibinfo{volume}{47},
  \bibinfo{number}{1} (\bibinfo{date}{Jan} \bibinfo{year}{2012}),
  \bibinfo{pages}{97–110}.
\newblock
\showISSN{0362-1340}
\urldef\tempurl%
\url{https://doi.org/10.1145/2103621.2103670}
\showDOI{\tempurl}


\bibitem[Canonne et~al\mbox{.}(2020)]%
        {Canonne:discrete:2020}
\bibfield{author}{\bibinfo{person}{Cl{\'e}ment~L. Canonne},
  \bibinfo{person}{Gautam Kamath}, {and} \bibinfo{person}{Thomas Steinke}.}
  \bibinfo{year}{2020}\natexlab{}.
\newblock \showarticletitle{The Discrete {{Gaussian}} for Differential
  Privacy}. In \bibinfo{booktitle}{\emph{Proceedings of the 34th
  {{International Conference}} on {{Neural Information Processing Systems}}}}
  \emph{(\bibinfo{series}{{{NIPS}} '20})}. \bibinfo{publisher}{Curran
  Associates Inc.}, \bibinfo{address}{Red Hook, NY, USA},
  \bibinfo{pages}{15676--15688}.
\newblock
\showISBNx{978-1-7138-2954-6}


\bibitem[Chatterjee et~al\mbox{.}(2020)]%
        {termination-book}
\bibfield{author}{\bibinfo{person}{Krishnendu Chatterjee},
  \bibinfo{person}{Hongfei Fu}, {and} \bibinfo{person}{Petr Novotn{\'{y}}}.}
  \bibinfo{year}{2020}\natexlab{}.
\newblock \showarticletitle{Termination Analysis of Probabilistic Programs with
  Martingales}.
\newblock In \bibinfo{booktitle}{\emph{Foundations of Probabilistic
  Programming}}, \bibfield{editor}{\bibinfo{person}{Gilles Barthe},
  \bibinfo{person}{Joost{-}Pieter Katoen}, {and} \bibinfo{person}{Alexandra
  Silva}} (Eds.). \bibinfo{publisher}{Cambridge University Press},
  \bibinfo{pages}{221--258}.
\newblock
\urldef\tempurl%
\url{https://doi.org/10.1017/9781108770750.008}
\showDOI{\tempurl}


\bibitem[Cowan et~al\mbox{.}(2024)]%
        {hands-on-dp}
\bibfield{author}{\bibinfo{person}{Ethan Cowan}, \bibinfo{person}{Michael
  Shoemate}, {and} \bibinfo{person}{Mayana Pereira}.}
  \bibinfo{year}{2024}\natexlab{}.
\newblock \bibinfo{booktitle}{\emph{Hands-on Differential Privacy: Introduction
  to the Theory and Practice Using {{OpenDP}}} (\bibinfo{edition}{first
  edition} ed.)}.
\newblock \bibinfo{publisher}{O'Reilly Media, Inc},
  \bibinfo{address}{Sebastopol, CA}.
\newblock
\showISBNx{978-1-4920-9774-7}
\showLCCN{QA76.9.A25 C675 2024}


\bibitem[{de Amorim} et~al\mbox{.}(2021)]%
        {deAmorim:Probabilistic:2021}
\bibfield{author}{\bibinfo{person}{Arthur~Azevedo {de Amorim}},
  \bibinfo{person}{Marco Gaboardi}, \bibinfo{person}{Justin Hsu}, {and}
  \bibinfo{person}{Shin-ya Katsumata}.} \bibinfo{year}{2021}\natexlab{}.
\newblock \showarticletitle{Probabilistic Relational Reasoning via Metrics}. In
  \bibinfo{booktitle}{\emph{Proceedings of the 34th {{Annual ACM}}/{{IEEE
  Symposium}} on {{Logic}} in {{Computer Science}}}}
  \emph{(\bibinfo{series}{{{LICS}} '19})}. \bibinfo{publisher}{IEEE Press},
  \bibinfo{address}{Vancouver, Canada}, \bibinfo{pages}{1--19}.
\newblock


\bibitem[{de Medeiros} et~al\mbox{.}(2025)]%
        {sampcert}
\bibfield{author}{\bibinfo{person}{Markus {de Medeiros}},
  \bibinfo{person}{Muhammad Naveed}, \bibinfo{person}{Tancr{\`e}de Lepoint},
  \bibinfo{person}{Temesghen Kahsai}, \bibinfo{person}{Tristan Ravitch},
  \bibinfo{person}{Stefan Zetzsche}, \bibinfo{person}{Anjali Joshi},
  \bibinfo{person}{Joseph Tassarotti}, \bibinfo{person}{Aws Albarghouthi},
  {and} \bibinfo{person}{Jean-Baptiste Tristan}.}
  \bibinfo{year}{2025}\natexlab{}.
\newblock \showarticletitle{Verified {{Foundations}} for {{Differential
  Privacy}}}.
\newblock \bibinfo{journal}{\emph{Artifact for Verified Foundations for
  Differential Privacy}} \bibinfo{volume}{9}, \bibinfo{number}{PLDI}
  (\bibinfo{date}{June} \bibinfo{year}{2025}),
  \bibinfo{pages}{191:1094--191:1118}.
\newblock
\urldef\tempurl%
\url{https://doi.org/10.1145/3729294}
\showDOI{\tempurl}


\bibitem[Dwork et~al\mbox{.}(2006)]%
        {Dwork:Calibrating:2006}
\bibfield{author}{\bibinfo{person}{Cynthia Dwork}, \bibinfo{person}{Frank
  McSherry}, \bibinfo{person}{Kobbi Nissim}, {and} \bibinfo{person}{Adam
  Smith}.} \bibinfo{year}{2006}\natexlab{}.
\newblock \showarticletitle{Calibrating {{Noise}} to {{Sensitivity}} in
  {{Private Data Analysis}}}. In \bibinfo{booktitle}{\emph{Theory of
  {{Cryptography}}}}, \bibfield{editor}{\bibinfo{person}{Shai Halevi} {and}
  \bibinfo{person}{Tal Rabin}} (Eds.). \bibinfo{publisher}{Springer},
  \bibinfo{address}{Berlin, Heidelberg}, \bibinfo{pages}{265--284}.
\newblock
\showISBNx{978-3-540-32732-5}
\urldef\tempurl%
\url{https://doi.org/10.1007/11681878_14}
\showDOI{\tempurl}


\bibitem[Dwork and Roth(2014)]%
        {Dwork:Algorithmic:2014}
\bibfield{author}{\bibinfo{person}{Cynthia Dwork} {and} \bibinfo{person}{Aaron
  Roth}.} \bibinfo{year}{2014}\natexlab{}.
\newblock \showarticletitle{The {{Algorithmic Foundations}} of {{Differential
  Privacy}}}.
\newblock \bibinfo{journal}{\emph{Foundations and Trends{\textregistered} in
  Theoretical Computer Science}} \bibinfo{volume}{9}, \bibinfo{number}{3--4}
  (\bibinfo{date}{Aug.} \bibinfo{year}{2014}), \bibinfo{pages}{211--407}.
\newblock
\showISSN{1551-305X, 1551-3068}
\urldef\tempurl%
\url{https://doi.org/10.1561/0400000042}
\showDOI{\tempurl}


\bibitem[Frumin et~al\mbox{.}(2021)]%
        {reloc}
\bibfield{author}{\bibinfo{person}{Dan Frumin}, \bibinfo{person}{Robbert
  Krebbers}, {and} \bibinfo{person}{Lars Birkedal}.}
  \bibinfo{year}{2021}\natexlab{}.
\newblock \showarticletitle{ReLoC Reloaded: {A} Mechanized Relational Logic for
  Fine-Grained Concurrency and Logical Atomicity}.
\newblock \bibinfo{journal}{\emph{Log. Methods Comput. Sci.}}
  \bibinfo{volume}{17}, \bibinfo{number}{3} (\bibinfo{year}{2021}).
\newblock
\urldef\tempurl%
\url{https://doi.org/10.46298/lmcs-17(3:9)2021}
\showDOI{\tempurl}


\bibitem[Gaboardi et~al\mbox{.}(2013)]%
        {dfuzz}
\bibfield{author}{\bibinfo{person}{Marco Gaboardi}, \bibinfo{person}{Andreas
  Haeberlen}, \bibinfo{person}{Justin Hsu}, \bibinfo{person}{Arjun Narayan},
  {and} \bibinfo{person}{Benjamin~C. Pierce}.} \bibinfo{year}{2013}\natexlab{}.
\newblock \showarticletitle{Linear {{Dependent Types}} for {{Differential
  Privacy}}}. In \bibinfo{booktitle}{\emph{Proceedings of the 40th Annual {{ACM
  SIGPLAN-SIGACT}} Symposium on {{Principles}} of Programming Languages}}
  \emph{(\bibinfo{series}{{{POPL}} '13})}. \bibinfo{publisher}{Association for
  Computing Machinery}, \bibinfo{address}{New York, NY, USA},
  \bibinfo{pages}{357--370}.
\newblock
\showISBNx{978-1-4503-1832-7}
\urldef\tempurl%
\url{https://doi.org/10.1145/2429069.2429113}
\showDOI{\tempurl}


\bibitem[Ghosh et~al\mbox{.}(2012)]%
        {Ghosh:Universally:2012}
\bibfield{author}{\bibinfo{person}{Arpita Ghosh}, \bibinfo{person}{Tim
  Roughgarden}, {and} \bibinfo{person}{Mukund Sundararajan}.}
  \bibinfo{year}{2012}\natexlab{}.
\newblock \showarticletitle{Universally {{Utility-maximizing Privacy
  Mechanisms}}}.
\newblock \bibinfo{journal}{\emph{SIAM J. Comput.}} \bibinfo{volume}{41},
  \bibinfo{number}{6} (\bibinfo{date}{Jan.} \bibinfo{year}{2012}),
  \bibinfo{pages}{1673--1693}.
\newblock
\showISSN{0097-5397}
\urldef\tempurl%
\url{https://doi.org/10.1137/09076828X}
\showDOI{\tempurl}


\bibitem[Gregersen et~al\mbox{.}(2024a)]%
        {caliper}
\bibfield{author}{\bibinfo{person}{Simon~Oddershede Gregersen},
  \bibinfo{person}{Alejandro Aguirre}, \bibinfo{person}{Philipp~G.
  Haselwarter}, \bibinfo{person}{Joseph Tassarotti}, {and}
  \bibinfo{person}{Lars Birkedal}.} \bibinfo{year}{2024}\natexlab{a}.
\newblock \showarticletitle{Almost-Sure Termination by Guarded Refinement}.
\newblock \bibinfo{journal}{\emph{CoRR}}  \bibinfo{volume}{abs/2404.08494}
  (\bibinfo{year}{2024}).
\newblock
\urldef\tempurl%
\url{https://doi.org/10.48550/ARXIV.2404.08494}
\showDOI{\tempurl}
\showeprint[arXiv]{2404.08494}


\bibitem[Gregersen et~al\mbox{.}(2024b)]%
        {clutch}
\bibfield{author}{\bibinfo{person}{Simon~Oddershede Gregersen},
  \bibinfo{person}{Alejandro Aguirre}, \bibinfo{person}{Philipp~G.
  Haselwarter}, \bibinfo{person}{Joseph Tassarotti}, {and}
  \bibinfo{person}{Lars Birkedal}.} \bibinfo{year}{2024}\natexlab{b}.
\newblock \showarticletitle{Asynchronous Probabilistic Couplings in
  Higher-Order Separation Logic}.
\newblock \bibinfo{journal}{\emph{Proc. {ACM} Program. Lang.}}
  \bibinfo{volume}{8}, \bibinfo{number}{{POPL}} (\bibinfo{year}{2024}),
  \bibinfo{pages}{753--784}.
\newblock
\urldef\tempurl%
\url{https://doi.org/10.1145/3632868}
\showDOI{\tempurl}


\bibitem[Haselwarter et~al\mbox{.}(2026a)]%
        {clutch-dp}
\bibfield{author}{\bibinfo{person}{Philipp~G. Haselwarter},
  \bibinfo{person}{Alejandro Aguirre}, \bibinfo{person}{Simon~Oddershede
  Gregersen}, \bibinfo{person}{Kwing~Hei Li}, \bibinfo{person}{Joseph
  Tassarotti}, {and} \bibinfo{person}{Lars Birkedal}.}
  \bibinfo{year}{2026}\natexlab{a}.
\newblock \showarticletitle{Modular Verification of Differential Privacy in
  Probabilistic Higher-Order Separation Logic}.
\newblock \bibinfo{journal}{\emph{Proc. ACM Program. Lang.}}
  \bibinfo{volume}{10}, \bibinfo{number}{PLDI}, Article
  \bibinfo{articleno}{233} (\bibinfo{date}{June} \bibinfo{year}{2026}).
\newblock
\urldef\tempurl%
\url{https://doi.org/10.1145/3808311}
\showDOI{\tempurl}


\bibitem[Haselwarter et~al\mbox{.}(2026b)]%
        {zenodo:clutch-dp}
\bibfield{author}{\bibinfo{person}{Philipp~G. Haselwarter},
  \bibinfo{person}{Alejandro Aguirre}, \bibinfo{person}{Simon~Oddershede
  Gregersen}, \bibinfo{person}{Kwing~Hei Li}, \bibinfo{person}{Joseph
  Tassarotti}, {and} \bibinfo{person}{Lars Birkedal}.}
  \bibinfo{year}{2026}\natexlab{b}.
\newblock \bibinfo{booktitle}{\emph{Modular Verification of Differential
  Privacy in Probabilistic Higher-Order Separation Logic - Artifact}}.
\newblock
\urldef\tempurl%
\url{https://doi.org/10.5281/zenodo.19089094}
\showDOI{\tempurl}


\bibitem[Haselwarter et~al\mbox{.}(2025)]%
        {approxis}
\bibfield{author}{\bibinfo{person}{Philipp~G. Haselwarter},
  \bibinfo{person}{Kwing~Hei Li}, \bibinfo{person}{Alejandro Aguirre},
  \bibinfo{person}{Simon~Oddershede Gregersen}, \bibinfo{person}{Joseph
  Tassarotti}, {and} \bibinfo{person}{Lars Birkedal}.}
  \bibinfo{year}{2025}\natexlab{}.
\newblock \showarticletitle{Approximate Relational Reasoning for Higher-Order
  Probabilistic Programs}.
\newblock \bibinfo{journal}{\emph{Proc. ACM Program. Lang.}}
  \bibinfo{volume}{9}, \bibinfo{number}{POPL}, Article \bibinfo{articleno}{41}
  (\bibinfo{date}{Jan.} \bibinfo{year}{2025}), \bibinfo{numpages}{31}~pages.
\newblock
\urldef\tempurl%
\url{https://doi.org/10.1145/3704877}
\showDOI{\tempurl}


\bibitem[Haselwarter et~al\mbox{.}(2024)]%
        {tachis}
\bibfield{author}{\bibinfo{person}{Philipp~G. Haselwarter},
  \bibinfo{person}{Kwing~Hei Li}, \bibinfo{person}{Markus de Medeiros},
  \bibinfo{person}{Simon~Oddershede Gregersen}, \bibinfo{person}{Alejandro
  Aguirre}, \bibinfo{person}{Joseph Tassarotti}, {and} \bibinfo{person}{Lars
  Birkedal}.} \bibinfo{year}{2024}\natexlab{}.
\newblock \showarticletitle{Tachis: Higher-Order Separation Logic with Credits
  for Expected Costs}.
\newblock \bibinfo{journal}{\emph{Proc. ACM Program. Lang.}}
  \bibinfo{volume}{8}, \bibinfo{number}{OOPSLA2}, Article
  \bibinfo{articleno}{313} (\bibinfo{date}{Oct.} \bibinfo{year}{2024}),
  \bibinfo{numpages}{30}~pages.
\newblock
\urldef\tempurl%
\url{https://doi.org/10.1145/3689753}
\showDOI{\tempurl}


\bibitem[Hsu(2017)]%
        {Hsu:Probabilistic:2017}
\bibfield{author}{\bibinfo{person}{Justin Hsu}.}
  \bibinfo{year}{2017}\natexlab{}.
\newblock \emph{\bibinfo{title}{Probabilistic {{Couplings}} for {{Probabilistic
  Reasoning}}}}.
\newblock \bibinfo{thesistype}{Ph.\,D. Dissertation}.
  \bibinfo{school}{University of Pennsylvania}.
\newblock
\showeprint[arxiv]{1710.09951}


\bibitem[Inusah and Kozubowski(2006)]%
        {Inusah:discrete:2006}
\bibfield{author}{\bibinfo{person}{Seidu Inusah} {and}
  \bibinfo{person}{Tomasz~J. Kozubowski}.} \bibinfo{year}{2006}\natexlab{}.
\newblock \showarticletitle{A Discrete Analogue of the {{Laplace}}
  Distribution}.
\newblock \bibinfo{journal}{\emph{Journal of Statistical Planning and
  Inference}} \bibinfo{volume}{136}, \bibinfo{number}{3} (\bibinfo{date}{March}
  \bibinfo{year}{2006}), \bibinfo{pages}{1090--1102}.
\newblock
\showISSN{0378-3758}
\urldef\tempurl%
\url{https://doi.org/10.1016/j.jspi.2004.08.014}
\showDOI{\tempurl}


\bibitem[Jung et~al\mbox{.}(2018)]%
        {irisjournal}
\bibfield{author}{\bibinfo{person}{Ralf Jung}, \bibinfo{person}{Robbert
  Krebbers}, \bibinfo{person}{Jacques{-}Henri Jourdan}, \bibinfo{person}{Ales
  Bizjak}, \bibinfo{person}{Lars Birkedal}, {and} \bibinfo{person}{Derek
  Dreyer}.} \bibinfo{year}{2018}\natexlab{}.
\newblock \showarticletitle{Iris from the ground up: {A} modular foundation for
  higher-order concurrent separation logic}.
\newblock \bibinfo{journal}{\emph{J. Funct. Program.}}  \bibinfo{volume}{28}
  (\bibinfo{year}{2018}), \bibinfo{pages}{e20}.
\newblock
\urldef\tempurl%
\url{https://doi.org/10.1017/S0956796818000151}
\showDOI{\tempurl}


\bibitem[Kostopoulou et~al\mbox{.}(2023)]%
        {turbo}
\bibfield{author}{\bibinfo{person}{Kelly Kostopoulou}, \bibinfo{person}{Pierre
  Tholoniat}, \bibinfo{person}{Asaf Cidon}, \bibinfo{person}{Roxana Geambasu},
  {and} \bibinfo{person}{Mathias L{\'e}cuyer}.}
  \bibinfo{year}{2023}\natexlab{}.
\newblock \showarticletitle{Turbo: {{Effective Caching}} in
  {{Differentially-Private Databases}}}. In
  \bibinfo{booktitle}{\emph{Proceedings of the 29th {{Symposium}} on
  {{Operating Systems Principles}}}} \emph{(\bibinfo{series}{{{SOSP}} '23})}.
  \bibinfo{publisher}{Association for Computing Machinery},
  \bibinfo{address}{New York, NY, USA}, \bibinfo{pages}{579--594}.
\newblock
\showISBNx{979-8-4007-0229-7}
\urldef\tempurl%
\url{https://doi.org/10.1145/3600006.3613174}
\showDOI{\tempurl}


\bibitem[Kotsogiannis et~al\mbox{.}(2019)]%
        {privatesql}
\bibfield{author}{\bibinfo{person}{Ios Kotsogiannis}, \bibinfo{person}{Yuchao
  Tao}, \bibinfo{person}{Xi He}, \bibinfo{person}{Maryam Fanaeepour},
  \bibinfo{person}{Ashwin Machanavajjhala}, \bibinfo{person}{Michael Hay},
  {and} \bibinfo{person}{Gerome Miklau}.} \bibinfo{year}{2019}\natexlab{}.
\newblock \showarticletitle{{{PrivateSQL}}: A Differentially Private {{SQL}}
  Query Engine}.
\newblock \bibinfo{journal}{\emph{Proc. VLDB Endow.}} \bibinfo{volume}{12},
  \bibinfo{number}{11} (\bibinfo{date}{July} \bibinfo{year}{2019}),
  \bibinfo{pages}{1371--1384}.
\newblock
\showISSN{2150-8097}
\urldef\tempurl%
\url{https://doi.org/10.14778/3342263.3342274}
\showDOI{\tempurl}


\bibitem[{Lobo-Vesga} et~al\mbox{.}(2020)]%
        {dpella}
\bibfield{author}{\bibinfo{person}{Elisabet {Lobo-Vesga}},
  \bibinfo{person}{Alejandro Russo}, {and} \bibinfo{person}{Marco Gaboardi}.}
  \bibinfo{year}{2020}\natexlab{}.
\newblock \showarticletitle{A {{Programming Framework}} for {{Differential
  Privacy}} with {{Accuracy Concentration Bounds}}}. In
  \bibinfo{booktitle}{\emph{2020 {{IEEE Symposium}} on {{Security}} and
  {{Privacy}} ({{SP}})}}. \bibinfo{pages}{411--428}.
\newblock
\showISSN{2375-1207}
\urldef\tempurl%
\url{https://doi.org/10.1109/SP40000.2020.00086}
\showDOI{\tempurl}


\bibitem[{Lobo-Vesga} et~al\mbox{.}(2024)]%
        {param-sens}
\bibfield{author}{\bibinfo{person}{Elisabet {Lobo-Vesga}},
  \bibinfo{person}{Alejandro Russo}, \bibinfo{person}{Marco Gaboardi}, {and}
  \bibinfo{person}{Carlos~Tom{\'e} Corti{\~n}as}.}
  \bibinfo{year}{2024}\natexlab{}.
\newblock \showarticletitle{Sensitivity by {{Parametricity}}}.
\newblock \bibinfo{journal}{\emph{Paper Artifact: Sensitivity by
  Parametricity}} \bibinfo{volume}{8}, \bibinfo{number}{OOPSLA2}
  (\bibinfo{date}{Oct.} \bibinfo{year}{2024}),
  \bibinfo{pages}{286:415--286:441}.
\newblock
\urldef\tempurl%
\url{https://doi.org/10.1145/3689726}
\showDOI{\tempurl}


\bibitem[Lohse and Garg(2024)]%
        {lohse2024iris}
\bibfield{author}{\bibinfo{person}{Janine Lohse} {and} \bibinfo{person}{Deepak
  Garg}.} \bibinfo{year}{2024}\natexlab{}.
\newblock \bibinfo{title}{An Iris for Expected Cost Analysis}.
\newblock
\newblock
\showeprint[arxiv]{2406.00884}~[cs.PL]


\bibitem[Lyu et~al\mbox{.}(2017)]%
        {Lyu:Understanding:2017}
\bibfield{author}{\bibinfo{person}{Min Lyu}, \bibinfo{person}{Dong Su}, {and}
  \bibinfo{person}{Ninghui Li}.} \bibinfo{year}{2017}\natexlab{}.
\newblock \showarticletitle{Understanding the Sparse Vector Technique for
  Differential Privacy}.
\newblock \bibinfo{journal}{\emph{Proc. VLDB Endow.}} \bibinfo{volume}{10},
  \bibinfo{number}{6} (\bibinfo{date}{Feb.} \bibinfo{year}{2017}),
  \bibinfo{pages}{637--648}.
\newblock
\showISSN{2150-8097}
\urldef\tempurl%
\url{https://doi.org/10.14778/3055330.3055331}
\showDOI{\tempurl}


\bibitem[Majumdar and Sathiyanarayana(2025)]%
        {termination-rules}
\bibfield{author}{\bibinfo{person}{Rupak Majumdar} {and} \bibinfo{person}{V.R.
  Sathiyanarayana}.} \bibinfo{year}{2025}\natexlab{}.
\newblock \showarticletitle{Sound and Complete Proof Rules for Probabilistic
  Termination}.
\newblock \bibinfo{journal}{\emph{Proc. ACM Program. Lang.}}
  \bibinfo{volume}{9}, \bibinfo{number}{POPL}, Article \bibinfo{articleno}{63}
  (\bibinfo{date}{Jan.} \bibinfo{year}{2025}), \bibinfo{numpages}{32}~pages.
\newblock
\urldef\tempurl%
\url{https://doi.org/10.1145/3704899}
\showDOI{\tempurl}


\bibitem[Mazmudar et~al\mbox{.}(2022)]%
        {cachedp}
\bibfield{author}{\bibinfo{person}{Miti Mazmudar}, \bibinfo{person}{Thomas
  Humphries}, \bibinfo{person}{Jiaxiang Liu}, \bibinfo{person}{Matthew Rafuse},
  {and} \bibinfo{person}{Xi He}.} \bibinfo{year}{2022}\natexlab{}.
\newblock \showarticletitle{Cache {{Me If You Can}}: {{Accuracy-Aware Inference
  Engine}} for {{Differentially Private Data Exploration}}}.
\newblock \bibinfo{journal}{\emph{Proc. VLDB Endow.}} \bibinfo{volume}{16},
  \bibinfo{number}{4} (\bibinfo{date}{Dec.} \bibinfo{year}{2022}),
  \bibinfo{pages}{574--586}.
\newblock
\showISSN{2150-8097}
\urldef\tempurl%
\url{https://doi.org/10.14778/3574245.3574246}
\showDOI{\tempurl}


\bibitem[Mironov(2012)]%
        {Mironov:significance:2012}
\bibfield{author}{\bibinfo{person}{Ilya Mironov}.}
  \bibinfo{year}{2012}\natexlab{}.
\newblock \showarticletitle{On Significance of the Least Significant Bits for
  Differential Privacy}. In \bibinfo{booktitle}{\emph{Proceedings of the 2012
  {{ACM}} Conference on {{Computer}} and Communications Security}}
  \emph{(\bibinfo{series}{{{CCS}} '12})}. \bibinfo{publisher}{Association for
  Computing Machinery}, \bibinfo{address}{New York, NY, USA},
  \bibinfo{pages}{650--661}.
\newblock
\showISBNx{978-1-4503-1651-4}
\urldef\tempurl%
\url{https://doi.org/10.1145/2382196.2382264}
\showDOI{\tempurl}


\bibitem[Near and Abuah(2021)]%
        {Near:Programming:2021}
\bibfield{author}{\bibinfo{person}{Joseph~P. Near} {and}
  \bibinfo{person}{Chik{\'e} Abuah}.} \bibinfo{year}{2021}\natexlab{}.
\newblock \bibinfo{booktitle}{\emph{Programming {{Differential}} {{Privacy}}}}.
  Vol.~\bibinfo{volume}{1}.
\newblock
\urldef\tempurl%
\url{https://programming-dp.com/}
\showURL{%
\tempurl}


\bibitem[Near et~al\mbox{.}(2019)]%
        {duet}
\bibfield{author}{\bibinfo{person}{Joseph~P. Near}, \bibinfo{person}{David
  Darais}, \bibinfo{person}{Chike Abuah}, \bibinfo{person}{Tim Stevens},
  \bibinfo{person}{Pranav Gaddamadugu}, \bibinfo{person}{Lun Wang},
  \bibinfo{person}{Neel Somani}, \bibinfo{person}{Mu Zhang},
  \bibinfo{person}{Nikhil Sharma}, \bibinfo{person}{Alex Shan}, {and}
  \bibinfo{person}{Dawn Song}.} \bibinfo{year}{2019}\natexlab{}.
\newblock \showarticletitle{Duet: An Expressive Higher-Order Language and
  Linear Type System for Statically Enforcing Differential Privacy}.
\newblock \bibinfo{journal}{\emph{Proc. ACM Program. Lang.}}
  \bibinfo{volume}{3}, \bibinfo{number}{OOPSLA} (\bibinfo{date}{Oct.}
  \bibinfo{year}{2019}).
\newblock
\urldef\tempurl%
\url{https://doi.org/10.1145/3360598}
\showDOI{\tempurl}


\bibitem[Peng et~al\mbox{.}(2013)]%
        {pioneer}
\bibfield{author}{\bibinfo{person}{Shangfu Peng}, \bibinfo{person}{Yin Yang},
  \bibinfo{person}{Zhenjie Zhang}, \bibinfo{person}{Marianne Winslett}, {and}
  \bibinfo{person}{Yong Yu}.} \bibinfo{year}{2013}\natexlab{}.
\newblock \showarticletitle{Query Optimization for Differentially Private Data
  Management Systems}. In \bibinfo{booktitle}{\emph{2013 {{IEEE}} 29th
  {{International Conference}} on {{Data Engineering}} ({{ICDE}})}}.
  \bibinfo{pages}{1093--1104}.
\newblock
\showISSN{1063-6382}
\urldef\tempurl%
\url{https://doi.org/10.1109/ICDE.2013.6544900}
\showDOI{\tempurl}


\bibitem[Reed and Pierce(2010)]%
        {fuzz}
\bibfield{author}{\bibinfo{person}{Jason Reed} {and}
  \bibinfo{person}{Benjamin~C. Pierce}.} \bibinfo{year}{2010}\natexlab{}.
\newblock \showarticletitle{Distance Makes the Types Grow Stronger: A Calculus
  for Differential Privacy}. In \bibinfo{booktitle}{\emph{Proceedings of the
  15th {{ACM SIGPLAN}} International Conference on {{Functional}} Programming}}
  \emph{(\bibinfo{series}{{{ICFP}} '10})}. \bibinfo{publisher}{Association for
  Computing Machinery}, \bibinfo{address}{New York, NY, USA},
  \bibinfo{pages}{157--168}.
\newblock
\showISBNx{978-1-60558-794-3}
\urldef\tempurl%
\url{https://doi.org/10.1145/1863543.1863568}
\showDOI{\tempurl}


\bibitem[Rogers et~al\mbox{.}(2016)]%
        {priv-filters}
\bibfield{author}{\bibinfo{person}{Ryan~M Rogers}, \bibinfo{person}{Aaron
  Roth}, \bibinfo{person}{Jonathan Ullman}, {and} \bibinfo{person}{Salil
  Vadhan}.} \bibinfo{year}{2016}\natexlab{}.
\newblock \showarticletitle{Privacy {{Odometers}} and {{Filters}}:
  {{Pay-as-you-Go Composition}}}. In \bibinfo{booktitle}{\emph{Advances in
  {{Neural Information Processing Systems}}}}, Vol.~\bibinfo{volume}{29}.
  \bibinfo{publisher}{Curran Associates, Inc.}
\newblock


\bibitem[Sato(2016)]%
        {Sato:Approximate:2016}
\bibfield{author}{\bibinfo{person}{Tetsuya Sato}.}
  \bibinfo{year}{2016}\natexlab{}.
\newblock \showarticletitle{Approximate Relational Hoare Logic for Continuous
  Random Samplings}. In \bibinfo{booktitle}{\emph{The Thirty-second Conference
  on the Mathematical Foundations of Programming Semantics, {MFPS} 2016}}.
\newblock
\urldef\tempurl%
\url{https://doi.org/10.1016/J.ENTCS.2016.09.043}
\showDOI{\tempurl}


\bibitem[Sato et~al\mbox{.}(2019)]%
        {Sato:Approximate:2019}
\bibfield{author}{\bibinfo{person}{Tetsuya Sato}, \bibinfo{person}{Gilles
  Barthe}, \bibinfo{person}{Marco Gaboardi}, \bibinfo{person}{Justin Hsu},
  {and} \bibinfo{person}{Shin-ya Katsumata}.} \bibinfo{year}{2019}\natexlab{}.
\newblock \showarticletitle{Approximate {{Span Liftings}}: {{Compositional
  Semantics}} for {{Relaxations}} of {{Differential Privacy}}}. In
  \bibinfo{booktitle}{\emph{2019 34th {{Annual ACM}}/{{IEEE Symposium}} on
  {{Logic}} in {{Computer Science}} ({{LICS}})}}. \bibinfo{pages}{1--14}.
\newblock
\urldef\tempurl%
\url{https://doi.org/10.1109/LICS.2019.8785668}
\showDOI{\tempurl}


\bibitem[Sato and Minamide(2025)]%
        {Sato:Formalization:2025}
\bibfield{author}{\bibinfo{person}{Tetsuya Sato} {and}
  \bibinfo{person}{Yasuhiko Minamide}.} \bibinfo{year}{2025}\natexlab{}.
\newblock \showarticletitle{Formalization of {{Differential Privacy}} in
  {{Isabelle}}/{{HOL}}}. In \bibinfo{booktitle}{\emph{Proceedings of the 14th
  {{ACM SIGPLAN International Conference}} on {{Certified Programs}} and
  {{Proofs}}}} \emph{(\bibinfo{series}{{{CPP}} '25})}.
  \bibinfo{publisher}{Association for Computing Machinery},
  \bibinfo{address}{New York, NY, USA}, \bibinfo{pages}{67--82}.
\newblock
\showISBNx{979-8-4007-1347-7}
\urldef\tempurl%
\url{https://doi.org/10.1145/3703595.3705875}
\showDOI{\tempurl}


\bibitem[Toro et~al\mbox{.}(2023)]%
        {jazz}
\bibfield{author}{\bibinfo{person}{Mat{\'i}as Toro}, \bibinfo{person}{David
  Darais}, \bibinfo{person}{Chike Abuah}, \bibinfo{person}{Joseph~P. Near},
  \bibinfo{person}{Dami{\'a}n {\'A}rquez}, \bibinfo{person}{Federico Olmedo},
  {and} \bibinfo{person}{{\'E}ric Tanter}.} \bibinfo{year}{2023}\natexlab{}.
\newblock \showarticletitle{Contextual {{Linear Types}} for {{Differential
  Privacy}}}.
\newblock \bibinfo{journal}{\emph{ACM Trans. Program. Lang. Syst.}}
  \bibinfo{volume}{45}, \bibinfo{number}{2} (\bibinfo{date}{May}
  \bibinfo{year}{2023}), \bibinfo{pages}{8:1--8:69}.
\newblock
\showISSN{0164-0925}
\urldef\tempurl%
\url{https://doi.org/10.1145/3589207}
\showDOI{\tempurl}


\bibitem[{Winograd-Cort} et~al\mbox{.}(2017)]%
        {afuzz}
\bibfield{author}{\bibinfo{person}{Daniel {Winograd-Cort}},
  \bibinfo{person}{Andreas Haeberlen}, \bibinfo{person}{Aaron Roth}, {and}
  \bibinfo{person}{Benjamin~C. Pierce}.} \bibinfo{year}{2017}\natexlab{}.
\newblock \showarticletitle{A Framework for Adaptive Differential Privacy}.
\newblock \bibinfo{journal}{\emph{Proc. ACM Program. Lang.}}
  \bibinfo{volume}{1}, \bibinfo{number}{ICFP} (\bibinfo{date}{Aug.}
  \bibinfo{year}{2017}), \bibinfo{pages}{10:1--10:29}.
\newblock
\urldef\tempurl%
\url{https://doi.org/10.1145/3110254}
\showDOI{\tempurl}


\bibitem[Zhang and Kifer(2017)]%
        {lightdp}
\bibfield{author}{\bibinfo{person}{Danfeng Zhang} {and} \bibinfo{person}{Daniel
  Kifer}.} \bibinfo{year}{2017}\natexlab{}.
\newblock \showarticletitle{{{LightDP}}: Towards Automating Differential
  Privacy Proofs}. In \bibinfo{booktitle}{\emph{Proceedings of the 44th {{ACM
  SIGPLAN Symposium}} on {{Principles}} of {{Programming Languages}}}}
  \emph{(\bibinfo{series}{{{POPL}} '17})}. \bibinfo{publisher}{Association for
  Computing Machinery}, \bibinfo{address}{New York, NY, USA},
  \bibinfo{pages}{888--901}.
\newblock
\showISBNx{978-1-4503-4660-3}
\urldef\tempurl%
\url{https://doi.org/10.1145/3009837.3009884}
\showDOI{\tempurl}


\bibitem[Zhang et~al\mbox{.}(2019)]%
        {fuzzi}
\bibfield{author}{\bibinfo{person}{Hengchu Zhang}, \bibinfo{person}{Edo Roth},
  \bibinfo{person}{Andreas Haeberlen}, \bibinfo{person}{Benjamin~C. Pierce},
  {and} \bibinfo{person}{Aaron Roth}.} \bibinfo{year}{2019}\natexlab{}.
\newblock \showarticletitle{Fuzzi: A Three-Level Logic for Differential
  Privacy}.
\newblock \bibinfo{journal}{\emph{Prototype Implementation of Fuzzi: A
  Three-Level Logic for Differential Privacy}} \bibinfo{volume}{3},
  \bibinfo{number}{ICFP} (\bibinfo{date}{July} \bibinfo{year}{2019}),
  \bibinfo{pages}{93:1--93:28}.
\newblock
\urldef\tempurl%
\url{https://doi.org/10.1145/3341697}
\showDOI{\tempurl}


\end{thebibliography}
